\documentclass[aps,pra,reprint,amsmath,amssymb,superscriptaddress,showpacs,nobibnotes,bibfootnote,showkeys]{revtex4-1}

\usepackage[table]{xcolor}
\usepackage{booktabs}
\usepackage{hhline}

\usepackage{graphicx}
\usepackage{dcolumn}
\usepackage{bm}
\usepackage{amsmath}
\usepackage{amssymb,amsthm}
\usepackage{bbm}
\usepackage{epsfig,color}
\usepackage[sans]{dsfont}

\graphicspath{ {images/} }

\definecolor{nblue}{rgb}{0.2,0.2,0.7}
\definecolor{ngreen}{rgb}{0.2,0.6,0.2}
\definecolor{nred}{rgb}{0.7,0.2,0.2}
\definecolor{nblack}{rgb}{0,0,0}

\newcommand{\ket}[1]{|#1\rangle}

\newcommand{\tr}{\text{tr}}

\renewcommand{\H}{\mathcal{H}}

\def\B{\mathcal{B}}

  \theoremstyle{definition}
  
  \theoremstyle{plain}
  
\theoremstyle{plain}
  
\theoremstyle{plain}

  \theoremstyle{plain}
  
  \theoremstyle{plain}

  \providecommand{\conjecturename}{Conjecture}
  \providecommand{\definitionname}{Definition}
  \providecommand{\lemmaname}{Lemma}
\providecommand{\corollaryname}{Corollary}
\providecommand{\theoremname}{Theorem}
\providecommand{\propositionname}{Proposition}

\def\m{\widetilde{m}}

\def\m{\widetilde{m}}

\def\H{\mathcal{H}}

\def\L{\mathrm{L}}
\def\U{\mathrm{U}}
\def\M{\mathrm{M}}
\def\S{\mathrm{S}}

\def\m{\Lambda_{\mathrm{TP}}^{(k)}}

\newcommand{\bk}[2]{\langle#1|#2\rangle}

\def\tr{\mbox{tr}}
\def\m{\widetilde{m}}
\def\bea{\begin{eqnarray}}
\def\eea{\end{eqnarray}}

\begin{document}

 \title{ Linking Entanglement Detection and State Tomography via Quantum $2$-Designs }

\author{Joonwoo Bae}
\affiliation{Department of Applied Mathematics, Hanyang University (ERICA), 55 Hanyangdaehak-ro, Ansan, Gyeonggi-do, 426-791, Korea}
\affiliation{ Freiburg Institute for Advanced Studies (FRIAS), Albert-Ludwigs University of Freiburg, Albertstrasse 19, 79104 Freiburg, Germany}
\author{Beatrix C. Hiesmayr}
\affiliation{University of Vienna, Faculty of Physics, Boltzmanngasse 5, 1090 Vienna, Austria}
\author{Daniel McNulty}
\affiliation{Department of Mathematics, Aberystwyth University, Aberystwyth, SY23 3BZ, UK}
\affiliation{Faculty of Informatics, Masaryk University, Brno, Czech Republic}

\date{March 7, 2018}

\begin{abstract}
We present an experimentally feasible and efficient method for detecting entangled states with measurements that extend naturally to a tomographically complete set. Our detection criterion is based on measurements from subsets of a quantum 2-design, e.g., mutually unbiased bases or symmetric informationally complete states, and has several advantages over standard entanglement witnesses. First, as more detectors in the measurement are applied, there is a higher chance of witnessing a larger set of entangled states, in such a way that the measurement setting converges to a complete setup for quantum state tomography. Secondly, our method is twice as effective as standard witnesses in the sense that both upper and lower bounds can be derived. Thirdly, the scheme can be readily applied to measurement-device-independent scenarios.

\end{abstract}



\maketitle


For quantum information applications, it is often more interesting to learn if multipartite quantum states are entangled than to identify quantum states themselves, e.g., \cite{intro1, intro2, intro3}. This is in fact what {\it direct detection of entanglement} executes, which aims to find if quantum states are entangled even before identifying quantum states. Entanglement witnesses (EWs) that work with individual measurements followed by post-processing of the outcomes \cite{terhal} provide an experimentally feasible approach for this purpose in general \cite{ewreport, ewreview}. Entanglement detection under less assumptions, for instance, when detectors are not trusted \cite{mdiew1, mdiew2,mdiew3} or dimensions are unknown \cite{semiew}, is of practical significance for cryptographic applications.

For the practical usefulness of entanglement detection, it is worth exploring the experimental resources. If {\it a priori} information about a quantum state is given, a set of EWs may be constructed accordingly and exploited for entanglement detection. With no {\it a priori} information multiple EWs may be required. One possible method is quantum state tomography (QST) which verifies a $d$-dimensional quantum state with $O(d^2)$ measurements. Then, theoretical tools such as positive maps \cite{pmap}, e.g. partial transpose, or numerical tests involving semidefinite programming \cite{sdp1, sdp2, sdp3} can be applied. For EWs, however, little is known about the minimal measurements for their realization. In fact, it may happen that repeating experiments for multiple EWs may be less cost effective than QST \cite{laflamme}, and quite possible that no useful information is obtained, neither for entanglement detection nor for quantum state identification. This raises questions on the usefulness of EWs, in particular when {\it a priori} information about a particular state is not available.

A useful experimental setup for entanglement detection may distinguish the largest collection of entangled states with as few measurements as possible. 
It is noteworthy that a tomographically complete measurement can ultimately identify a quantum state so that theoretical tools may completely determine whether it is entangled or separable. From a practical point of view, it would be therefore highly desirable that measurements for entanglement detection are constructive, i.e., they can be extended to a tomographically complete set by augmenting more detectors.

In this work we establish a feasible and practical framework of entanglement detection by applying a subset of measurements taken from a quantum $2$-design, namely mutually unbiased bases (MUBs)~\cite{ref:schwinger} and symmetric informationally complete states (SICs) \cite{ref:renes}. The connections between entanglement detection, MUBs, and quantum $2$-designs have first been explored in Refs. \cite{ref:spengler, ref:bae}, and subsequent results were found in, e.g.~\cite{ref:chen, ref:li, ref:graydon}.
 Let us emphasize here that the detection via MUBs is in some cases more powerful than the Peres-Horodecki criterion since also bound entangled states, those mixed entangled states from which no entanglement can be distilled, are detected. Furthermore, measurement setups with MUBs are very experimentally friendly, indeed the MUB criterion~\cite{ref:spengler} resulted in the first experimental demonstration of bipartite bound entanglement~\cite{ref:hiesmayr2, ref:hiesmayr1}, predicted in 1998~\cite{ref:Horodecki}. Here we present a unifying approach to these connections with a three-fold advantage. First, by using incomplete sets of MUBs and SICs, the entanglement detection scheme then extends naturally to an optimal reconstruction of the quantum state \cite{mubd2,scott2006}: once direct detection of entanglement fails, additional detectors are applied in the measurement scheme to distinguish a larger set of entangled states, and can be ultimately utilised to find its separability via state tomography. This demonstrates in a natural framework that larger sets of detectors are more useful for distinguishing entangled states. Next, our results have {\it twice} the efficiency of standard EWs, in the sense that both a lower and upper bound for separable states exist, whereas EWs have only the zero-valued lower bound. Finally, the scheme can be readily applied to a measurement-device-independent (MDI) scenario for which the assumptions on the detectors are relaxed. This can be achieved by converting the measurement into the preparation of a quantum $2$-design.

Let us begin with a brief summary on the implementation of EWs in practice. EWs correspond to observables that have non-negative expectation values for all separable states as well as negative values for some entangled states. They can be factorized into local observables in general, which are then decomposed by positive-operator-valued-measure (POVM) elements. A witness $W$ can be written with POVMs denoted by $\{M_{i}^{(X)} \}$ for party $X=A,B$, where the measurement is complete, i.e., $\sum_{i} M_{i}^{(X)} = \mathrm{I}_{X}$ where $\mathrm{I}_X$ denotes the identity operator on system $X$, as
\bea
W = \sum_{i} c_i~ M_i,~~\mathrm{where}~~ M_i = M_{i}^{(A)} \otimes M_{i}^{(B)}, \label{eq:1}
\eea
with constants $\{ c_i \}$. In implementation, a POVM element can be realized by projective measurements with ancillary systems, see e.g., \cite{naimark}. For a state $\rho$, the probabilities $\mathrm{Pr}[M_i | \rho] = \tr[\rho M_i]$ are estimated experimentally by the detectors $\{ M_i \}$. Then, the expectation value of $W$ for a state $\rho$ is obtained by computing the linear combination, $\sum_{i}c_i \mathrm{Pr} [M_i | \rho ]$, which equals $\tr[W\rho]$.

Although the factorization with local measurements in Eq.~(\ref{eq:1}) is not necessary to realize EWs, it provides a natural framework for converting standard EWs to the MDI scenario that closes all loopholes arising from detectors. In such a scenario two parties Alice and Bob, who want to learn if an unknown quantum state $\rho_{AB}$ is entangled, prepare a set of quantum states, after which a measurement is performed by untrusted parties. A standard witness in Eq. (\ref{eq:1}) can be used to construct an MDI-EW as follows,
\bea
W_{\mathrm{MDI}} = \sum_{i} c_i~ M_{i}^{(A)\top } \otimes M_{i}^{(B)\top}, \label{eq:2}
\eea
where the transpose $\top$ is performed in a chosen basis of $\H_Y$ for $Y=A,B$ \cite{mdiew2}. The separable decomposition in Eq. (\ref{eq:2}) shows which quantum states the two parties must prepare, $\{ \widetilde{M}_{i}^{(A) } \}$ and $\{ \widetilde{M}_{i}^{(B) } \}$, where $\widetilde{M}_{i}^{( Y) } = M_{i}^{(Y) } / \tr[M_{i}^{(Y) }]$ correspond to the quantum states.

Let us reiterate that EWs with local measurements in Eq. (\ref{eq:1}) are readily converted to their counterparts in an MDI scenario, where entangled states are detected with less assumptions. We also note that, to the best of our knowledge, there is no general and systematic way of finding the factorization with a minimal number of local measurements. The decomposition with a minimal number of POVMs is essential, as mentioned, to take the advantage of EWs which can detect entangled states without QST.

We now introduce a particular set of POVMs called a quantum $2$-design. A set of quantum states $\{|\psi_i\rangle \}_k$ in a $d$-dimensional Hilbert space, $|\psi_i \rangle \in \H_d$, or their corresponding rank-one operators, is called a quantum $2$-design if the average value of any second order polynomial over the set $\{|\psi_i\rangle \}_k$ is equal to the average $f(\psi)$ over all normalized states given a suitable measure, such as the Haar measure. This holds true if and only if the average of $| \psi_i \rangle \langle  \psi_i |^{\otimes 2}$ over the entire $2$-design is proportional to the symmetric projection onto $\H_d \otimes \H_d$. A complete set of $(d+1)$ MUBs, and a SIC-POVM containing $d^2$ elements, are both quantum $2$-designs. In fact, the existence of $(d+1)$ MUBs and $d^2$ SIC states in all dimensions have been long-standing open problems in quantum information theory \cite{ref:openmub, ref:opensic}. For instance, complete sets of MUBs are known to exist in prime-power dimensions~\cite{mubd1, mubd2, mubd3, mubd4, mubd5, mubd6,mubd7} but have not been found in in any other composite dimension. For example, when $d=6$, it is conjectured that only $3$ MUBs exist~\cite{mubdp1, mubdp2, mubdp3, mubdp4,mubdp5}, but no proof exists. While it is conjectured that a SIC-POVM exists for any $d$, the largest dimension for which an example has been found is $d=323$ \cite{scott2017}.

Let $\B_k = \{|b_{i}^{k} \rangle \}_{i=1}^{d} $ denote a set of MUBs in the Hilbert space $\H_d$, and let $S_d = \{ |s_k\rangle \}_{k=1}^{d^2}$ denote a SIC-POVM in the same Hilbert space. The two sets satisfy the equations
\bea
|\langle b_{i}^{l} | b_{j}^{k}\rangle |^2 = d^{-1},~\mathrm{and}~ |\langle s_k | s_l \rangle |^2 = (d+1)^{-1}, ~~\label{mubsic}
\eea
respectively, for all $k\neq l$. It is well known that a full set of $(d+1)$ MUBs and a SIC-POVM are tomographically complete: measurements from either set determine a quantum state uniquely. Furthermore, the sets are both optimal and simple for QST, in that they minimize the error of the estimated statistics while at the same time having exceptionally simple state reconstruction formulas \cite{mubd2,scott2006}. Note that both MUBs and SIC-POVMs are experimentally feasible, and have been implemented for the purpose of QST. A recent demonstration has been given in \cite{ref:sicexp1}.

\begin{figure}
\begin{center}
\includegraphics[scale=0.14]{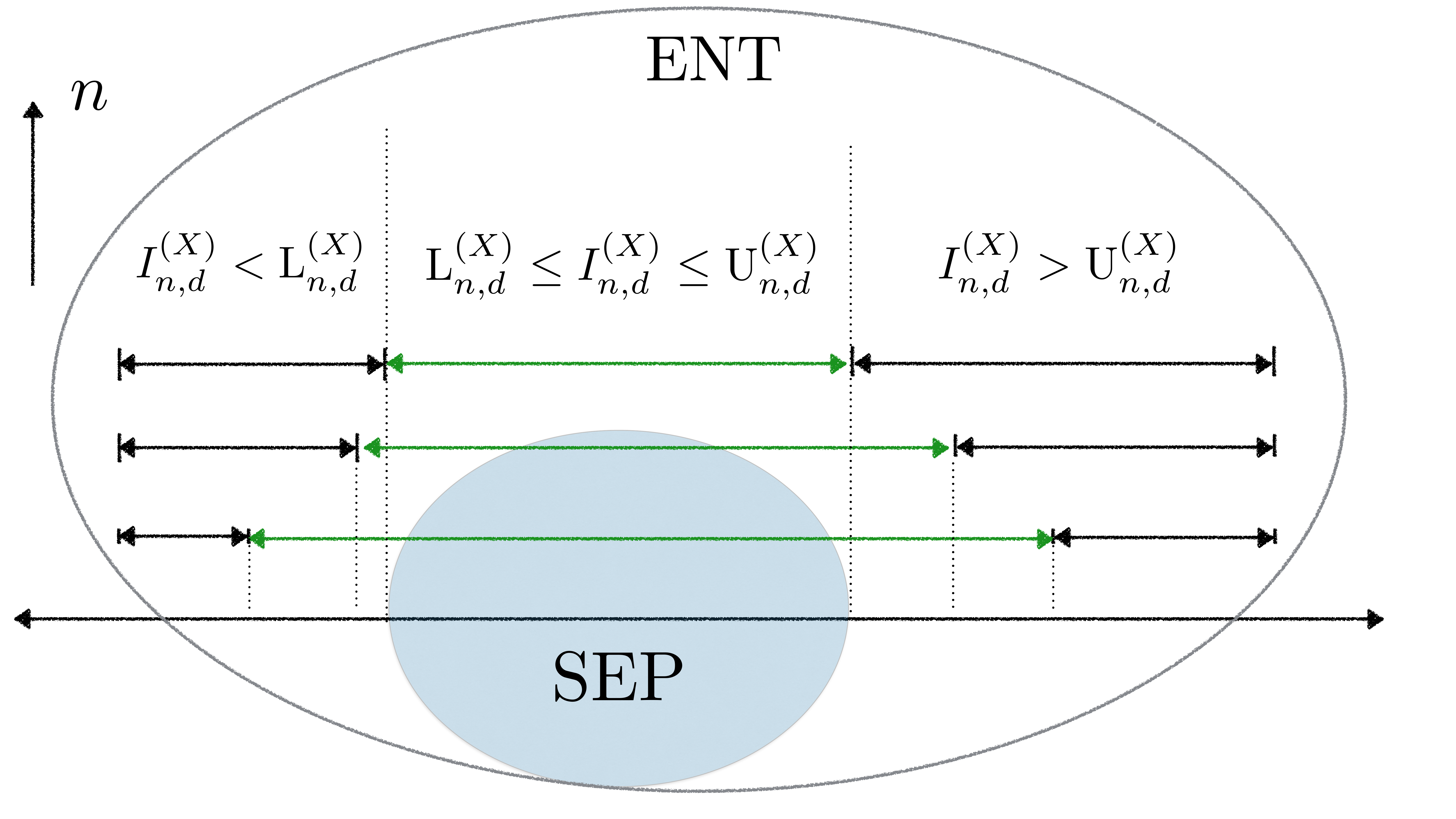}
\caption{ Our strategy for detecting entangled states via MUBs and SICs is illustrated, where $X = \M, \S$ and $n=m,\m$, see inequalities in Eqs. (\ref{eq:mubinq}) and (\ref{eq:sicinq}) satisfied by all separable states. Violation of the bounds implies detection of entangled states. Once the measurement outcomes are collected, they are exploited {\it twice} to find if the upper or lower bound is violated, in which case entangled states are detected. }
\label{fig:illustration}
\end{center}
\end{figure}

We now consider tomographically incomplete sets of MUBs and SICs for detecting entangled states. We denote by $I_{m,d}^{( \mathrm{M})}$ and $I_{\m,d}^{( \mathrm{S})}$ the collections of probabilities when the measurements are applied in MUBs and SICs, respectively,
\bea
I_{m,d}^{(\mathrm{M})} (\rho: \{ \B_k\}_{k=1}^{m}) & = & \sum_{k=1}^{m} \sum_{i=1}^{d} \mathrm{Pr} ( i , i | \B_k, \B_k),~~     \label{eq:imub} \\
I_{\m,d}^{(\mathrm{S})} (\rho:S_{\m} ) & = & \sum_{j=1}^{\m}  \mathrm{Pr} ( j , j | S_{\m} ,S_{\m}),~~  \label{eq:isic}
\eea
where $S_{\m}$ denotes a collection of $\m$ states out of $d^2$ SICs, and $ \mathrm{Pr} ( \alpha, \beta | A,B)$ the probability of obtaining outcome $(\alpha, \beta)$ given a measurement in $A$ and $B$. To be explicit, for state $\rho$, $\mathrm{Pr} ( i , i | \B_k, \B_k) = \tr[ |b_{i}^{k} \rangle \langle b_{i}^{k} | \otimes  |b_{i}^{k} \rangle \langle b_{i}^{k} |  ~ \rho ]$ and $\mathrm{Pr} ( j,j | S_{\m}, S_{\m}) = \tr[ | s_{j} \rangle \langle s_{j}  | \otimes | s_{j} \rangle \langle s_{j}  | ~ \rho ]$ \cite{footnote}. These probabilities can be obtained simply by preparing local measurements in MUBs or SICs. Note that we have $m\leq d+1 $ and $\m\leq d^2$, where the equality corresponds to cases that the measurement setting is tomographically complete. Then, from the measurements one can construct the quantum state for which one can apply all theoretically known criteria to detect entanglement.

Since the set of all separable states forms a convex set, the quantities $I_{m,d}^{(\mathrm{M})}$ and $I_{\m,d}^{(\mathrm{S})}$ as defined in Eqs. (\ref{eq:imub}) and (\ref{eq:isic}) have both nontrivial upper and lower bounds satisfied by all separable states. In what follows, the bounds for selections of $m$ MUBs and $\m$ SICs are explicitly presented. We minimize and maximize each of the bounds with respect to the set of MUBs and SICs, e.g., minimizing (maximizing) the lower bound over all MUBs gives $\L_{m,d}^{- (\mathrm{M})} $ ( $\L_{m,d}^{+(\mathrm{M})} $). The former (latter) gives a bound which is independent (dependent) of the choice of MUBs. Consequently, $\L_{m,d}^{+(\mathrm{M})} $ detects a larger set of entangled states but only applies for a certain collection of MUBs.


%

\setlength{\arrayrulewidth}{0.5pt}
\begin{table}[h!]
\begin{center}
\begin{tabular}{||c|c|c|c|c||c|c|c||}
\hhline{|t=====:t:===t|}
&\multicolumn{4}{c||}{Lower Bounds}&\multicolumn{3}{c||}{Upper Bounds}\\
\hhline{||-|-|-|-|-||-|-|-||}
&$\cellcolor{blue!50}d=2$&$\cellcolor{red!50}d=3$&\multicolumn{2}{c||}{\cellcolor{green!50}$d=4$}& \cellcolor{blue!50}$d=2$&\cellcolor{red!50}$d=3$&\cellcolor{green!50}$d=4$\\
\hhline{||-|-|-|-|-||-|-|-||}
$m$&$\vphantom{\biggl\lbrace} \L_{m,2}^{(\mathrm{M})}$ ~&~ $\L_{m,3}^{(\mathrm{M})}$ ~&~ $\L_{m,4}^{- (\mathrm{M})}$ ~&~ $\L_{m,4}^{+ (\mathrm{M})}$  ~&~ $\U_{m,2}^{(\mathrm{M})}$ ~&~ $\U_{m,3}^{(\mathrm{M})}$
~&~  $\U_{m,4}^{(\mathrm{M})}$    \\
\hhline{|:=====||===:|}
$2$  & 1/2& 0.211& 0  & 0  & 3/2 & 4/3 & 5/4  \\ \hhline{||-|-|-|-|-|-|-|-||}
$3$  &  1& 1/2& 1/4 & 1/2   & 2    & 5/3 &  6/4 \\ \hhline{||-|-|-|-|-|-|-|-||}
$4$ & & 1& 1/2  & 1/2          &   &    2  & 7/4 \\ \hhline{||-|-|-|-|-|-|-|-||}
 $5$ & & & 1  & 1  &  &  &  2  \\
\hhline{|b:=====:b:===:b|}
\end{tabular}
\end{center}
\caption{Lower and upper bounds on MUBs, $\L_{m,d}^{\pm (\mathrm{M})}$ and $\U_{m,d}^{(\M)}$, see Eqs. (\ref{eq:mubl1}), (\ref{eq:mubl2}) and (\ref{eq:mubu}), are summarized for $m$ MUBs in $\H = \mathbb{C}^d$, for $d=2,3,4$. For $d=2,3$, different full sets of MUBs are unitarily equivalent, hence we have $\L_{m,d}^{+ (\mathrm{M})} = \L_{m,d}^{ - (\mathrm{M})}$. }
\label{tab:MUB}
\end{table}

%

When the measurements are taken from a set of MUBs, the minimal and maximal lower bounds, $\L_{m,d}^{- (\mathrm{M})} $ and $\L_{m,d}^{+(\mathrm{M})} $, respectively, are given by
\bea
\L_{m,d}^{- (\mathrm{M})} &=& \min_{\{ \B_{k} \}_{k=1}^m} \min_{\sigma_{\mathrm{sep}}} ~ I_{m,d}^{(\mathrm{M})} (\sigma_{\mathrm{sep}} :  \{ \B_{k} \}_{k=1}^m ), ~~\label{eq:mubl1}   \\
\L_{m,d}^{+ (\mathrm{M})} &=& \max_{ \{ \B_{k} \}_{k=1}^m } \min_{\sigma_{\mathrm{sep}}} ~  I_{m,d}^{(\mathrm{M})} (\sigma_{\mathrm{sep}} : \{ \B_{k} \}_{k=1}^m ), ~~ \label{eq:mubl2}
\eea
where the optimisation is taken over all separable states $\sigma_{\text{sep}}$ and all possible collections of $m$ MUBs, $\{ \B_{k} \}_{k=1}^m$, that exist in dimension $d$. It is clear that $\L_{m,d}^{+ (\mathrm{M})} \geq \L_{m,d}^{- (\mathrm{M})} $, and the gap between the bounds is due to different sets of $m$ MUBs having different overlaps with the set of separable states.

Unfortunately, we do not find a systematic and general method of obtaining these bounds but had to consider all possible sets of $m$ MUBs minimizing $I_{m,d}^{ (\M)}$ over all separable states. In Table \ref{tab:MUB}, lower bounds are shown for $d=2,3,4$, which are obtained analytically. It turns out that $\L_{m,d}^{- (\mathrm{M})} = \L_{m,d}^{ + (\mathrm{M})}$ for $d=2,3$, but for $d=4$ we found $\L_{m,4}^{- (\mathrm{M})} \geq \L_{m,4}^{ + (\mathrm{M})}$. The difference here is due to the existence of an infinite family of 3 MUBs in $d=4$, resulting in unitarily inequivalent triples. The triple which gives $ \L_{m,4}^{ - (\mathrm{M})}=1/4$ is the only extendible set of 3 MUBs, in the sense that no other triple extends to a complete set of 5 MUBs. For $d=2,3,$ all subsets of $m$ MUBs are equivalent and extendible.

In Ref.~\cite{ref:spengler}, it has been shown that the upper bound does not depend on selections of MUBs, and is given by
\bea
\U_{m,d}^{(\mathrm{M})} & = & \max_{\sigma_{\mathrm{sep}}} ~ I_{m,d}^{(\mathrm{M})} ( \sigma_{\mathrm{sep}}: \{ \B_{k}\}_{ k=1}^{m} ) =  1 + \frac{m-1}{d}, ~~~\label{eq:mubu}
\eea
for any $m$ MUBs $\{ \B_{k } \}_{k =1}^{m}$.  Note that in the case of a quantum $2$-design with $m=d+1$, the upper bound satisfies $\U_{d+1,d}^{(\mathrm{M})} =2$, which is independent of the dimension $d$. Notice also that by removing a single basis from $I_{m,d}^{( \mathrm{M})}$ the upper bound decreased uniformly by $1/d$, i.e.,
\bea
\U_{m+1 ,d }^{(\mathrm{M})} - \U_{m ,d }^{(\mathrm{M})} ~= ~d^{-1} \nonumber
\eea
for all $m$ MUBs.

In our first main result, using Table~\ref{tab:MUB} and Eq.~(\ref{eq:mubu}), we can construct the inequalities with optimization over $m$ MUBs in Eq.~(\ref{eq:imub}) as
\bea
\L_{m,d}^{-(\M)} \leq I_{m,d}^{(\M)} (\sigma_{\mathrm{sep}} ) \leq \U_{m,d}^{(\M)}\,, \label{eq:mubinq}
\eea
that are satisfied by all separable states in $\H_d \otimes \H_d$. A quantum state must be entangled if it violates one of the inequalities above, see also Fig. \ref{fig:illustration}. It is also worth mentioning that these inequalities detect bound entangled states when $m=d+1$, as shown in \cite{ref:hiesmayr2, ref:hiesmayr1}.

\begin{table}[h!]
\begin{center}
\begin{tabular}{||c | c | c | c || c | c | c||}
\hhline{|t====:t:===t|}
&\multicolumn{3}{c||}{Lower Bounds}&\multicolumn{3}{c||}{Upper Bounds}\\
\hhline{||-|-|-|-||-|-|-||}
&\cellcolor{blue!50}$d=2$&\multicolumn{2}{c||}{\cellcolor{red!50}$d=3$}& \cellcolor{blue!50}$d=2$&\multicolumn{2}{c||}{\cellcolor{red!50}$d=3$}\\
\hhline{||-|-|-|-||-|-|-||}
 $\m~ $   &  $\vphantom{\biggl\lbrace}  ~\L_{\m,2}^{(\mathrm{S})}$   ~ & ~$ \L_{\m,3}^{ - (\mathrm{S} ) } $ ~ &~ $\L_{\m, 3}^{+(\mathrm{S})} $  ~&  ~ $\U_{\m ,2}^{(\mathrm{S})}$  ~ &~ $ \U_{\m, 3}^{+ ( \mathrm{S} ) } $ ~ & ~  $\U_{\m, 3}^{- (\mathrm{S})}$ \\
\hhline{|:====||===:|}
 3  ~&  0 &  0      & 0 &      1.244      & 1.254 & 9/8  \\ \hhline{||----||---||}
 4  ~&  4/15 & 0      & 0 &    4/3    &1.400 &  1.25  \\ \hhline{||----||---||}
 5  ~&  2/3 & 0      & 0   &    4/3    &1.463  &  1.400 \\\hhline{||----||---||}
 6  ~&        & 0      & 0.112&     & 3/2   &  1.482  \\ \hhline{||----||---||}
 7  ~&       & 3/20 & 3/20     &   & 3/2      &   3/2 \\ \hhline{||----||---||}
 8  ~&       & 3/8   & 3/8    &     & 3/2      &    3/2 \\ \hhline{||----||---||}
 9  ~&       & 3/4   &  3/4  &      & 3/2      &    3/2 \\
\hhline{|b====:b:===b|}
\end{tabular}
\end{center}
\caption{The lower and upper bounds via SICs, $\L_{\m,d}^{\pm (\S)}$ and $\U_{\m,d}^{\pm (\S)}$, are shown for $d=2,3$. For $d=2$ there is only one SIC-POVM while for $d=3$ we use the Hesse SIC defined in the Appendix. Note that $\L_{\m,2}^{ + (\mathrm{S})} = \L_{\m,2}^{ - (\mathrm{S})}$ and $\U_{\m,2}^{ + (\mathrm{S})} = \U_{\m,2}^{ - (\mathrm{S})}$. In contrast to MUBs, we find that $\U_{\m,d}^{ + (\mathrm{S})} \geq \U_{\m,d}^{ - (\mathrm{S})}$.  }
\label{tab:SICs}
\end{table}

In a similar way, lower and upper bounds for SICs are denoted as follows, with $g=\pm$, and $\mathrm{opt^{+}} = \max$ and $\mathrm{opt^{-}} = \min$,
\bea
\L_{\m,d}^{g ~(\mathrm{S })} &=& \mathrm{opt^g}_{ S_{\m}\subseteq S_{d^2} } \min_{\sigma_{\mathrm{sep}}} ~ I_{\m,d}^{(\mathrm{S })} (\sigma_{\mathrm{sep}} :  S_{\m}) ~\mathrm{and}~~\label{eq:sicl}  \\
\U_{\m,d}^{g~ (\mathrm{S })} &=& \mathrm{opt^g}_{ S_{\m}\subseteq S_{d^2} } \max_{\sigma_{\mathrm{sep}}} ~I_{\m,d}^{(\mathrm{S })} (\sigma_{\mathrm{sep}} :  S_{\m}), ~~\label{eq:sicu}
\eea
where $S_{\m}$ is a set of $\m$ SICs. Then, the full set of SICs is denoted by $S_{d^2}$. Again, we do not find a systematic and general method of computing upper and lower bounds. However, having explored all possible subsets of SICs in $d=2,3$, for a given SIC-POVM, we present these bounds in Table \ref{tab:SICs}. Suboptimal bounds for $d=4$ are also presented in the Appendix. We observe that $\U_{\m,d}^{ + (\mathrm{S})} \geq \U_{\m,d}^{ - (\mathrm{S})}$, i.e., differences in the subsets of SICs give rise to the gap between these upper bounds. Therefore, the inequalities which are satisfied by all separable states are constructed in our second main result as
\bea
\L_{\m,d}^{-(\S)} \leq I_{\m,d}^{(\S)} (\sigma_{\mathrm{sep}} ) \leq \U_{\m,d}^{+(\S)}\,, \label{eq:sicinq}
\eea
where $\L_{\m,d}^{-(\S)}$ and $\U_{\m,d}^{+(\S)}$ are found in Table \ref{tab:SICs}. Even tighter inequalities with $\L_{\m,d}^{+(\S)}$ and $\U_{\m,d}^{-(\S)}$ can be derived by specifying the corresponding subset of $\m$ SICs. We note that for large $\m$ the upper bounds become independent of the choice of SICs, e.g., $U^{+(S)}_{\m,3}=U^{-(S)}_{\m,3}=3/2$ for $\m=7,8,9$.


\begin{figure}
\includegraphics[scale=0.13]{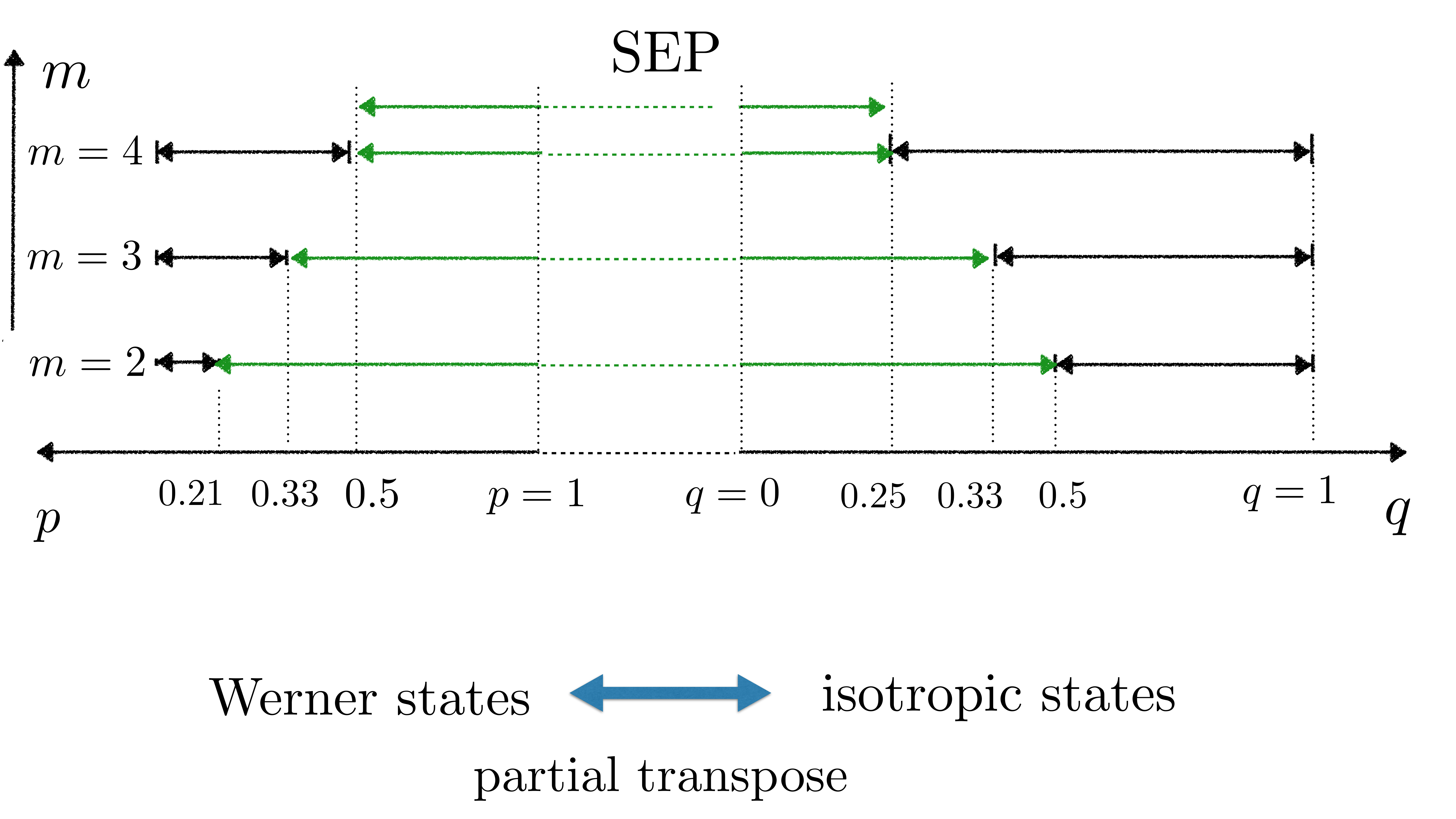}
\caption{The inequalities $I_{2,3}^{(\mathrm{M})}$,  $I_{3,3}^{(\mathrm{M })}$,  and $I_{4,3}^{(\mathrm{M })}$ are applied to detect entangled states. Once $I_{m,d}^{}(\mathrm{M})$ for unknown quantum states is obtained, it can be utilized twice for entanglement detection with both upper and lower bounds. E.g., the upper bounds are violated by entangled isotropic states and the lower bounds by entangled Werner states.}
\label{fig:picmub}
\end{figure}

While these inequalities have been obtained by extensively considering all sets of MUBs and SICs, analytic expressions for the upper and lower bounds can be derived for a quantum $2$-design,
\bea
1\leq I_{d+1,d}^{(\M)}(\sigma_{\mathrm{sep}})\leq 2,~~\frac{d}{d+1} \leq I_{d^2,d}^{(\S)}(\sigma_{\mathrm{sep}})\leq \frac{2d}{d+1},~~~~ \label{eq:inqd}
\eea
as shown in the Appendix. The upper bounds to $I_{d+1,d}^{(\M)}$ and $I_{d^2,d}^{(\S)}$ are proven in Refs. \cite{ref:spengler} and \cite{ref:li}, respectively. Lower bounds are shown in Ref. \cite{ref:bae} and later in Ref. \cite{ref:graydon}. As mentioned earlier, when the full measurement set of a quantum 2-design is used, it is more efficient to exploit the measurements for QST, and use theoretical tools to solve the separability problem that is known to be $NP$-hard.

To illustrate the effectiveness of the inequalities in Eqs. (\ref{eq:mubinq}) and (\ref{eq:sicinq}), consider the isotropic and Werner states,
\bea
&& \mathrm{Werner ~state}: ~\rho_{\mathrm{W}} ( p )  = p\; \widetilde{\Pi }_{\mathrm{sym} } + (1-p)\; \widetilde{\Pi }_{\mathrm{asym}}  \label{eq:werner} \\
&& \mathrm{isotropic ~ state}: ~\rho_{\mathrm{iso} } (q)  = q | \Phi^+ \rangle \langle \Phi^{+} | + (1-q ) \mathbbm{1}_d \otimes \mathbbm{1}_d~~~~~~\label{eq:iso}
\eea
where $\widetilde{\Pi}_{\mathrm{sym}}$ and $\widetilde{\Pi}_{\mathrm{asym}}$ denote the normalized projections onto the symmetric and anti-symmetric subspaces, respectively, and $\mathbbm{1}_d = \mathbbm{1}/d$, the normalized identity operator in dimension $d$. It is known that $\rho_{\mathrm{W}}$ is entangled iff $p<1/2$ and $\rho_{\mathrm{iso}}$ iff $q>(d+1)^{-1}$. In Fig. \ref{fig:picmub}, the capability of entanglement detection with $I_{m,3}^{(\M)}$ is shown for $m=2,3,4$. The capability of entanglement detection via SICs is given in the Appendix.

Due to the linearity of Eqs. (\ref{eq:imub}) and (\ref{eq:isic}), with respect to the state $\rho$, one may expect that the inequalities in Eq. (\ref{eq:inqd}) are closely connected to standard EWs. Here we point out the equivalence between the lower bounds in Eq. (\ref{eq:inqd}) and the partial transpose criterion, by considering the so-called structural physical approximation (SPA) \cite{intro1}. For recent reviews on the SPA see \cite{intro2, intro3}, as well as the Appendix for further details. The Choi-Jamiolkowski (CJ) operator for the transpose map corresponds to an EW, denoted by $W$, i.e., $\tr[\sigma_{\mathrm{sep} }W] \geq 0$, and $\tr[\rho W]<0$ for some entangled states $\rho$ which include the entangled Werner states in Eq. (\ref{eq:werner}). By applying the SPA to the transpose map, the resulting CJ operator denoted by $\widetilde{W}$ is given by $\widetilde{W} = \widetilde{\Pi}_{\mathrm{sym}}$. The condition $\tr[\sigma_{\mathrm{sep} }W] \geq 0$ then translates to $\tr[\sigma_{\mathrm{sep} } \widetilde{W}] \geq  [d(d+1)]^{-1}$, see Ref. \cite{ref:bae}, which is equivalent to the lower bounds in Eq. (\ref{eq:inqd}).

Finally, we can see that $I_{m,d}^{(\M)} (\rho) = \tr[ W_{m,d}^{(\M)}  \rho]$ and $I_{\m,d}^{(\S)} (\rho) = \tr[ W_{\m,d}^{(\S)}  \rho]$ are readily converted for entanglement detection in a MDI scenario where,
\bea
W_{m,d}^{( \M)} (\{ \B_{k}\}_{k=1}^m  ) &=& \sum_{k=1}^{m} \sum_{i=1}^d | b_{i}^{k} \rangle \langle  b_{i}^{k}| \otimes | b_{i}^{k} \rangle \langle  b_{i}^{k}|, \nonumber  \\
W_{\m,d}^{( \S)} (S_{\m}  ) &=& \sum_{j = 1}^{\m}  | s_{j}  \rangle \langle  s_{j} | \otimes | s_{j}  \rangle \langle  s_{j} |. \nonumber
\eea
As described in Eq. (\ref{eq:2}), both $I_{m,d}^{(\M)}$ and $I_{\m,d}^{(\S)}$ can be obtained in an MDI manner with $W_{m,d}^{( \M) \top} (\{ \B_{k}\}_{k=1}^m  )$ and $W_{\m,d}^{( \S) \top} (S_{\m}  ) $, respectively, by preparing the set of quantum states $\{ \B_{ k} \}_{k=1}^m$ and $S_{\m}$ instead of measurements in these bases. Note also that this provides both upper and lower MDI bounds as opposed to standard MDI-EWs.


To conclude, let us recall the problem addressed at the outset. How do we learn efficiently if an unknown quantum state is entangled, with a measurement that is tomographically incomplete? We also assume that, for practical purposes, the setup is constructive in that it can be easily extended to that of QST. While EWs are useful for direct detection of entanglement, it is highly non-trivial to compare and connect their measurements to those which are useful for QST. However, this is a crucial requirement when experimentalists decide whether to perform direct detection of entanglement or ultimately add more detectors to identify the separability problem via state reconstruction. Our results achieve this objective with a measurement setup which can detect entangled states with cost effective measurements, and which extend naturally to the tomographically complete setup of a quantum 2-design which allows for optimal state reconstruction. Furthermore, they offer {\it double} the efficiency of standard and non-linear EWs by providing both upper and lower bounds. One consequence of our analysis is that certain sets of MUBs are more `useful' for entanglement detection than others. For instance, in dimension $d=4$, the set of 3 MUBs which extends to a complete set provides the minimal (weakest) lower bound and therefore detects a smaller set of entangled states than unextendible MUBs. Thus, one might expect that unextendible MUBs are more useful in other dimensions too. We also note that the results can be generalized to weighted $2$-designs \cite{roy}, which would allow for entanglement detection and QST in dimensions where the existence of MUBs and SICs is not yet known.

We envisage directions in entanglement detection beyond standard EWs and towards related problems in quantum information theory. While we have already shown some links between standard EWs and the MUB-inequality~(\ref{eq:mubinq}) and the SIC-inequality~(\ref{eq:sicinq}), we expect further connections to also hold true.  For example, recently it has been shown that MUBs can be used to construct positive but not completely positive maps, which lead to a class of EWs \cite{darek}. Further relations in this direction may reveal additional capabilities of EWs at an even deeper level. It would also be interesting to consider nonlinearity, e.g., in Ref. \cite{nl2}, to improve the inequalities. We also hope that the presented framework of entanglement detection may offer insightful hints towards a solution of the existence problem for MUBs and SICs from an entanglement perspective \cite{ref:openmub, ref:opensic}. In addition, MUBs and SICs have quite recently been generalized by relaxing the rank-$1$ condition to so-called mutually unbiased measurements (MUMs) and symmetric informationally complete measurements (SIMs), which exist in all finite dimensions \cite{ref:kalevgourmub, ref:kalevgoursic}. Both MUMs and SIMs, as well as other similar measurements, could be applied to our framework in similar ways, leading to more experimentally feasible entanglement detection methods in arbitrary dimensions.

\section*{Acknowledgement}

J.B. is supported by the Institute for Information \& communications Technology Promotion(IITP) grant funded by the Korea government(MSIP) (R0190-17-2028), National Research Foundation of Korea (NRF-2017R1E1A1A03069961), the KIST Institutional Program (2E26680-17-P025), and the People Programme (Marie Curie Actions) of the European Union Seventh Framework Programme (FP7/2007- 2013) under REA grant agreement N. 609305. B.C.H gratefully acknowledges the Austrian Science Fund FWF-P26783. D.M. has received funding from the European Union's Horizon 2020 research and innovation programme under the Marie Sk\l{}odowska-Curie grant agreement No 663830.

\appendix
\section*{Appendices}
\vspace{0.5cm}

\section{Quantum $2$-Designs, MUBs and SICs}
\label{sec:2design}

In these appendices we review known results on quantum $2$-designs, mutually unbiased bases (MUBs), symmetric informationally complete measurements (SICs), and entanglement witnesses (EWs). The main results are presented, including a derivation of the lower and upper bounds for inequalities which detect entangled states via collections of MUBs and SICs. We analyse the capability of our criterion, and show that as we apply more measurements, i.e., as the number of MUBs and SICs increase, the criterion detects larger sets of entangled states. When we apply a quantum 2-design, i.e., a full set of $(d+1)$ MUBs or $d^2$ SICs, the inequalities provide a necessary and sufficient condition for the separability of a certain class of quantum states, namely the symmetric states. We also show for quantum 2-designs how our detection criterion is related to EWs.

Let us begin with a discussion on quantum $2$-designs, also known as complex projective $2$-designs, and two well known examples, a complete set of $(d+1)$ MUBs and a SIC-POVM consisting of $d^2$ elements. An ensemble of $n$ normalized $d$-dimensional vectors $\mathcal{D}=\{\ket{\psi_k}\}\subseteq \mathbb{C}^d$ is a quantum $2$-design if the average value of any second order polynomial $f(\psi)$ over the set $\mathcal{D}$ is identical to the average of $f(\psi)$ over the unitarily invariant Haar distribution of unit vectors $\ket{\psi}\in\mathbb{C}^d$. To be precise, $f(\psi)$ is a homogenous polynomial of degree two in the coefficients of $\ket{\psi}$ and of degree two in the complex conjugates of these coefficients. In other words, $\mathcal{D}$ is a quantum $2$-design if it has the first two moments equal to those of the Haar distribution. It can be shown that such an ensemble of vectors  is a quantum $2$-design if and only if
\bea\label{eq:design_condition}
\frac{1}{n} \sum_{i=1}^n|\psi_i \rangle \langle \psi_i |^{\otimes 2} = \frac{2}{d(d+1)}\Pi_{\text{sym}}, \label{eq:2design}
\eea
where $\Pi_{\text{sym}}$ is the projector onto the symmetric subspace of $\mathbb{C}^d\otimes \mathbb{C}^d$.

We write the symmetric and anti-symmetric projectors,
\bea
\Pi_{\text{sym}} = \frac{1}{2} (\mathbbm{1}_d\otimes \mathbbm{1}_d + \Pi),~\mathrm{and}~ \Pi_{\text{asym}} = \frac{1}{2}(\mathbbm{1}_d\otimes \mathbbm{1}_d - \Pi)\nonumber
\eea
respectively, where $\mathbbm{1}_d$ denotes the identity operator in $d$-dimensional Hilbert space, and $\Pi$ corresponds to the permutation operator in $\B( \mathbb{C}^d\otimes \mathbb{C}^d)$. Note the useful relation that $\Pi^{\Gamma} = d |\Phi^{+} \rangle \langle \Phi^{+} |$, with $\Gamma$ the partial transpose and $|\Phi^{+} \rangle=\frac{1}{\sqrt{d}}\sum_{i=1}^{d} |ii\rangle$ the maximally entangled state.


Well known examples of quantum $2$-designs are complete sets of $(d+1)$ MUBs and a SIC-POVM. Let $\mathcal{B}_k=\{ |b^{k}_i \rangle\}_{i=1}^d$ denote an orthonormal basis of the space $\mathbb{C}^d$. $\B_{k}$ and $\B_{l}$ are called mutually unbiased if it holds that for all $i,j$, $| \langle b_{i}^{k}  | b_{j}^{\ell} \rangle |^2 = d^{-1}$. SIC states are a set of normalized vectors $\{ |s_k\rangle \}_{k=1}^{\m}$ in $\mathbb{C}^d$ satisfying the relation $| \langle s_{k}   | s_{ l}  \rangle |^2 = (d+1)^{-1}$ for all $k\neq l$. The SIC states form a SIC-POVM when $\m=d^2$. Suppose that for a $d$-dimensional Hilbert space, there exist $(d+1)$ MUBs and $d^2$ SIC states. Then, it holds that
\bea
\widetilde{ \Pi}_{\mathrm{sym}} & = & \frac{1}{d(d+1)}\sum_{k=1}^{d+1} \sum_{i=1}^d |b_{i}^{k} \rangle \langle b_{i}^{k}|^{\otimes 2}= \frac{1}{d^2} \sum_{k=1}^{d^2} |s_k\rangle \langle s_k|^{\otimes 2}\nonumber
\eea
where $\widetilde{\Pi}_{\mathrm{sym}}$ denotes the normalized projection onto the symmetric subspace, $\widetilde{\Pi}_{\mathrm{sym}}  = 2 [ d(d+1)]^{-1} \Pi_{\mathrm{sym}} $.

Note that the existence of a complete set of MUBs and a SIC-POVM has been a long-standing open problem in quantum information theory and is related to several other unsolved problems in mathematics such as orthogonal decompositions of Lie algebras. It is conjectured that there exist $(d+1)$ MUBs if and only if the dimension $d$ is a prime-power, while a set of $d^2$ SICs is conjectured to exist for all $d$~\cite{ref:zauner}. So far, it is known that complete sets of MUBs exist in all prime-power dimensions~\cite{mubd1, mubd2, mubd3, mubd4, mubd5, mubd6,mubd7}, while only significantly smaller sets have been found in other composite dimensions. In particular, for dimension $d=6$, numerical calculations suggest that there exist only $3$ MUBs~\cite{mubdp1, mubdp2, mubdp3}. On the other hand, it is known that a SIC-POVM exists in all dimensions $d \leq 323$~\cite{scott2017,ref:fuchs}.


\section{Detecting Entangled States Using MUBs and SICs}
\label{sec:de}

Let us now consider incomplete sets of MUBs and SICs for entanglement detection. We will formulate the inequalities in terms of probabilities, having both upper and lower bounds, which are satisfied by all separable states. Since the structure of MUBs and SICs is not fully understood, it is a non-trivial task to derive these bounds. For instance, in certain dimensions $d$, different equivalence classes of MUBs exist, and the bounds can often depend on the choice of a particular class. Furthermore, the bounds do not appear to have a simple analytical expression, behaving differently as the dimension changes. In the following, we will first consider entanglement detection with measurements corresponding to MUBs, and then apply similar techniques to derive bounds for SICs. Finally, we show the relationship between quantum $2$-designs and EWs.


\begin{figure}
\begin{center}
\includegraphics[scale=0.13]{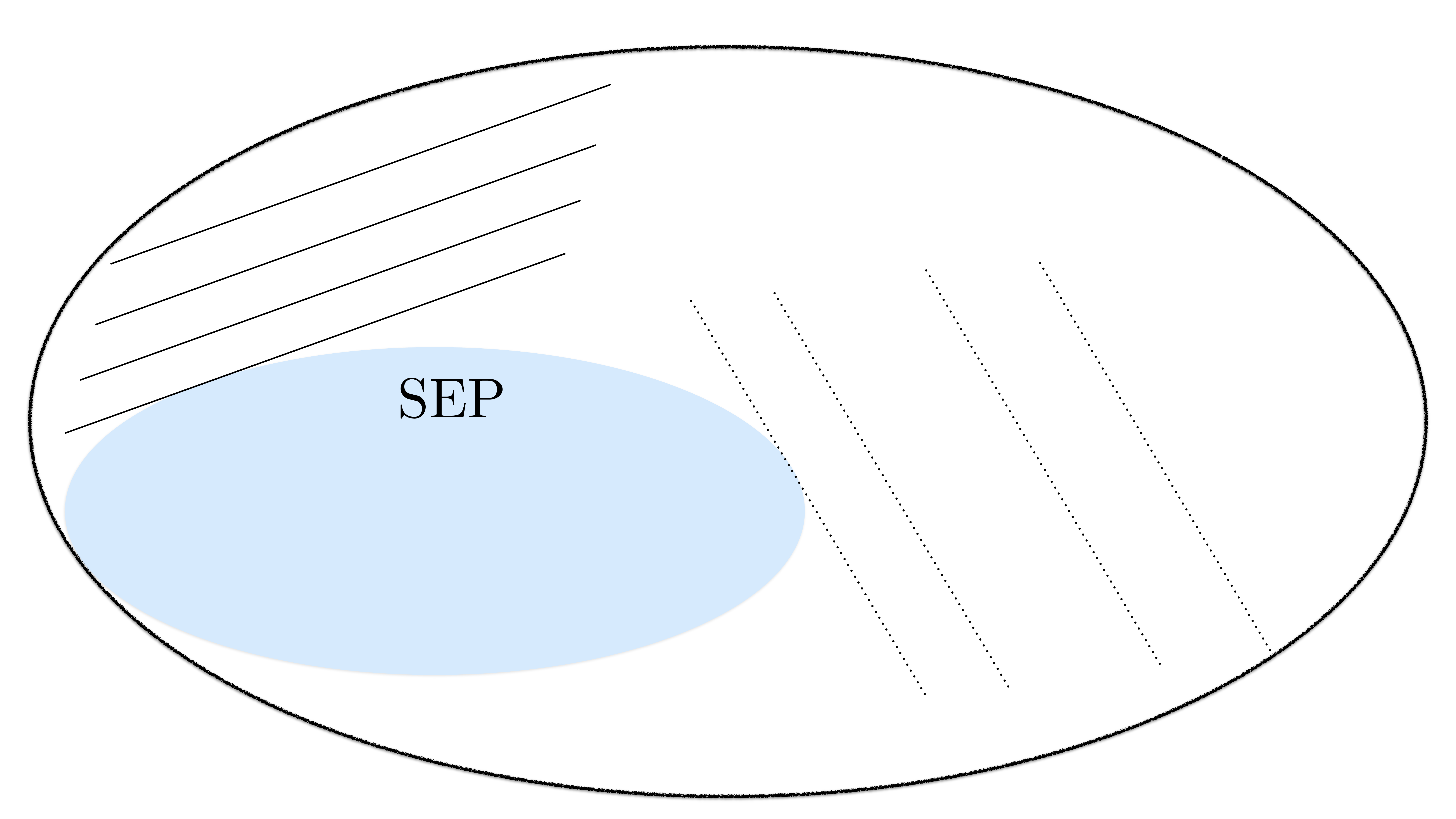}
\caption{Entanglement detection via MUBs and SICs is illustrated, as shown by the inequalities presented in Eqs. (\ref{eq:mublu}) and (\ref{eq:siclu}). Since both the upper and lower bounds are linear with respect to quantum states, they correspond to distinct hyperplanes that separate some entangled states from separable ones. Violation of either bound detect entangled states.}
\label{fig:illustration1}
\end{center}
\end{figure}

We denote by $I_{m,d}^{(\mathrm{M})}$ and $I_{\m,d}^{\mathrm{(S)}}$, collections of probabilities when measurements are applied from sets of MUBs and SICs, respectively. For measurements of a set of $m$ MUBs, $\{ \B_{k}\}_{k=1}^m$ in $\mathbb{C}^d$, or a set $S_{\m}\subseteq\mathbb{C}^d$ of $\m$ SIC states from a SIC-POVM, applied to each subsystem of a $(d\times d)$ bipartite state $\rho$, these quantities are defined as,
\bea
I_{m,d}^{(\mathrm{M})} (\rho : \{\B_{k}\}_{k=1}^{m} )  & = & \sum_{k=1}^m \sum_{i=1}^d \mathrm{Pr} (i,i | \B_k, \B_k) ,\label{eq:mubc} \\
I_{\m,d}^{\mathrm{(S)}} (\rho:S_{\m}) & = & \sum_{j=1}^{\m} \mathrm{Pr} (j,j | S_{\m}, S_{\m}), \label{eq:sicc}
\eea
where we have $\mathrm{Pr} (i,i | \B_k, \B_k) = \tr[ |b_{i}^{k} \rangle \langle b_{i}^{k}| \otimes |b_{i}^{k} \rangle \langle  b_{i}^{k}|  ~\rho]$ and $\mathrm{Pr} ( j,j | S_{\m}, S_{\m}) = \tr[ |s_j  \rangle \langle  s_j  | \otimes | s_j  \rangle \langle  s_j  | ~\rho]$. We now derive upper and lower bounds for the quantities $I_{m,d}^{(\mathrm{M})}$ and $I_{\m,d}^{(\mathrm{S})}$, which hold true for all separable states.

\subsection{Lower and upper bounds of $I_{m,d}^{(\mathrm{M})}$}
\label{subsec:mub}

Let $\L_{m,d}^{(\mathrm{M})}$ and $\U_{m,d}^{(\mathrm{M})}$ denote the upper and lower bounds of $I_{m,d}^{(\mathrm{M})}$, respectively, for a set of $m$ MUBs, $\{ \B_{k}\}_{k=1}^m$, with $1< m \leq d+1$. We calculate these quantities by minimizing and maximizing over all separable states such that
\bea
\L_{m,d}^{(\mathrm{M} )} (\{ \B_{k}\}_{k =1}^m) & = & \min_{\sigma_{\mathrm{sep}}} ~ I_{m,d}^{(\mathrm{M})} (\sigma_{\mathrm{sep}} : \{ \B_{k}\}_{k=1}^m), \label{mublg} \\
\U_{m,d}^{(\mathrm{M} )} (\{ \B_{k}\}_{k=1}^m) & = & \max_{\sigma_{\mathrm{sep}}} ~ I_{m,d}^{(\mathrm{M})} (\sigma_{\mathrm{sep}} : \{ \B_{k}\}_{k=1}^m). \label{mubug}
\eea
For certain dimensions $d$, there exists inequivalent sets of $m$ MUBs, up to unitary transformations. For instance, some sets extend to $(d+1)$ MUBs while others are unextendible \cite{grassl17}. Thus, the bounds above may also have a dependence on the choice of MUBs, and hence we also classify these additional bounds as follows,
\bea
\L_{m,d}^{- ( \mathrm{M}) } & = & \min_{\{ \B_k\}_{k=1}^m } \L_{m,d}^{(\mathrm{M} )} (\{ \B_{k}\}_{ k =1}^m), \label{eq:mubl-}  \\
\L_{m,d}^{+ (\mathrm{M} )} & = & \max_{\{ \B_k\}_{k=1}^m }\L_{m,d}^{(\mathrm{M} )} (\{ \B_{k}\}_{ k =1}^m),  \label{eq:mubl+}\\
\U_{m,d}^{- ( \mathrm{M} )} & = & \min_{\{ \B_k\}_{k=1}^m } \U_{m,d}^{(\mathrm{M} )} (\{ \B_{k }\}_{k=1}^m),  \label{eq:mubu-} \\
\U_{m,d}^{+ (\mathrm{M} )} & = & \max_{\{ \B_k\}_{k=1}^m }\U_{m,d}^{(\mathrm{M} )} (\{ \B_{k }\}_{k=1}^m), \label{eq:mubu+}
\eea
where the minimum and maximum are taken over all possible collections of $m$ MUBs, $\{\mathcal{B}_k\}_{k=1}^m$, that exist in dimension $d$. Note that for $d\leq 5$, all sets of MUBs are known \cite{brierley10,haagerup}. However, for $d\geq6$, the complete classification of MUBs remains an open problem, even for prime-power dimensions, hence such an optimization is currently not possible in large dimensions.

It then follows we have the bounds,
\bea
\L_{m,d}^{ - (\mathrm{M })} ~\leq ~\L_{m,d}^{ +(\mathrm{M })} ~\leq  ~I_{m,d}^{(\mathrm{M})} (\sigma_{\mathrm{sep}}) ~\leq ~ \U_{m,d}^{(\mathrm{M})}\,,\label{eq:mublu}
\eea
that are satisfied by all separable states. We will show in the next section that the upper bound $\U_{m,d}^{(\mathrm{M})}$ is independent of the choice of MUBs, i.e., $\U_{m,d}^{(\mathrm{M})}=\U_{m,d}^{\pm(\mathrm{M})}$.
The tighter lower bound, $\L_{m,d}^{ +(\mathrm{M })}$, applies only for a particular set of MUBs, i.e., the set which maximizes $\L_{m,d}^{(\mathrm{M})}$ in Eq. (\ref{eq:mubl+}). We also note that the minimal lower bound $\L_{m,d}^{ - (\mathrm{M })}$ applies for any choice of $m$ MUBs. Thus, entangled states are detected by observing violations of $\L_{m,d}^{ - (\mathrm{M })}$ and $\U_{m,d}^{(\mathrm{M})}$ regardless of the choice of MUBs.

\subsubsection{Upper bound $\U_{m,d}^{(\mathrm{M})}$ }

In Ref. \cite{ref:spengler} the upper bound has no dependence on the selection of $m$ MUBs and it is shown that
\bea
\U_{m,d}^{(\mathrm{M})} := \U_{m,d}^{\pm (\mathrm{M})} = 1 + \frac{m-1}{d}. \label{eq:mubuAppendix}
\eea
We note that for $m=d+1$, i.e., the quantum $2$-design case, the upper bound is given by $\U_{d+1,d}^{(\mathrm{M})}=2$ and is clearly independent of the dimension $d$.

\begin{table}
\begin{center}
\begin{tabular}{c | c | c| c}
\hline \hline
~$m$ ~& ~$\vphantom{\biggl\lbrace}~ \U_{m,2}^{(\mathrm{M})}$~ &~ $\U_{m,3}^{(\mathrm{M})}$ ~& ~$\U_{m,4}^{(\mathrm{M})}$ \\ \hline \hline
$2$  & 3/2&  4/3     & 5/4   \\ \hline
$3$  &  2  &  5/3     & 6/4 \\ \hline
$4$  &      &   2       & 7/4  \\ \hline
 $5$ &      &            &  2  \\
\hline \hline
\end{tabular}
\end{center}
\caption{Upper bounds $\U_{m,d}^{(\mathrm{M})}$ in Eq. (\ref{eq:mubuAppendix}) are summarized for $m$ MUBs in $\mathbb{C}^d$, for $d=2,3,4$. As the number of MUBs decreases from $m$ to $m-1$, the upper bound is reduced uniformly by $1/d$. }
\label{tab:MUBs_upperbound}
\end{table}

We also observe that removing a single basis from the set of $m$ MUBs decreases the upper bound uniformly by $1/d$, i.e.,
\bea
\U_{m+1,d}^{(\mathrm{M })} - \U_{m,d}^{(\mathrm{M })} = d^{-1}, \label{eq:mubupper}
\eea
and the bound is not influenced by which basis is subtracted from the set of MUBs. The bounds for $d=2,3,4$ are summarized in Table \ref{tab:MUBs_upperbound}.

\subsubsection{Lower bound $\L_{m,d}^{(\mathrm{M})}$ }

For the lower bounds of $I_{m,d}^{(\mathrm{M})}$, the minimization and maximization of Eqs. (\ref{eq:mubl-}) and (\ref{eq:mubl+}) over all MUBs do not coincide in general, i.e., $\L_{m,d}^{ - (\mathrm{M })}\leq \L_{m,d}^{ +(\mathrm{M })}$. Let us first consider the minimization in Eq. (\ref{mublg}) for $m$ MUBs, $\{ \B_k\}_{k=1}^m$. Recall that a separable state can be decomposed by a convex combination of product states. This means that it suffices to consider the minimization over only product states, as follows,
\bea
\L_{m,d}^{(\mathrm{M})} (  \{ \B_k \}_{k=1}^m)  :=  \min_{|e\rangle,|f\rangle}  \sum_{k=1}^m\sum_{i=1}^{d}|\bk{b^k_i}{e}|^2|\bk{b_i^k}{f}|^2, ~~~~~~\label{eq:mubl}
\eea
where $\mathcal{B}_k=\{\ket{b^k_i}\}_{i=1}^{d}$, and unit vectors $\ket{e},\ket{f}\in\mathbb{C}^d$. To obtain the minimal and maximal bounds in Eqs. (\ref{eq:mubl-}) and (\ref{eq:mubl+}), the optimization must run over all selections of $m$ MUBs that exist. We do not yet have a systematic method of finding optimal sets of $m$ MUBs that give the tight and minimal lower bounds $\L_{m,d}^{+(\mathrm{M})}$ and $\L_{m,d}^{-(\mathrm{M})}$. In what follows, we derive these bounds for dimensions $d=2,3,4$.

The property of equivalence classes of MUBs, up to unitary or anti-unitary transformations, is useful to simplify the numerical optimizations in Eqs. (\ref{eq:mubl-}) and (\ref{eq:mubl+}). We call a set of $m$ MUBs, $\{ \mathcal{B}_k\}_{k=1}^m$, equivalent to another set of $m$ MUBs, $\{ \mathcal{B}_{k}^{'} \}_{k=1}^m$, denoted by
\bea
\{ \B_{k}  \}_{k=1}^m \sim \{ \B_{k}^{'} \}_{k=1}^m,\nonumber
\eea
if there exists a unitary or anti-unitary transformation, denoted by $V$, such that $\mathcal{B}_k = V \mathcal{B}_{k}^{'} V^{\dagger}$ for $k=1,\dots, m$. Note that equivalent sets of $m$ MUBs give the same values for $I_{m,d}^{(\mathrm{M})}$:
\bea
&& \{ \B_{k}  \}_{k=1}^m \sim \{ \B_{k}^{'} \}_{k=1}^m \Rightarrow \nonumber\\
&& I_{m,d}^{(\mathrm{M})} (\sigma_{\mathrm{sep}} : \{ \B_k \}_{k=1}^m ) =  I_{m,d}^{(\mathrm{M})} (\sigma_{\mathrm{sep}} : \{ \B_{k}^{'} \}_{k=1}^m )\,. ~~~~~\label{eq:equiv}
\eea
The converse, however, does not hold true in general. It therefore suffices to consider distinct equivalence classes in the optimization of Eqs. (\ref{eq:mubl-}) and (\ref{eq:mubl+}).

It turns out that, for dimensions $d=2,3$, all sets of $m$ MUBs with $m\leq d+1$ are equivalent. In these low dimensions, the optimization in Eqs. (\ref{eq:mubl-}) and (\ref{eq:mubl+}) is not necessary, and hence, for any $m$ MUBs,
\bea
\L_{m,d}^{ (\mathrm{M})}  := \L_{m,d}^{(\mathrm{M })} (  \{ \B_k \}_{k=1}^m )=\L_{m,d}^{\pm(\mathrm{M})}, \nonumber
\eea
In Table \ref{tab:MUBs_lowerbound}, these lower bounds are listed as $\L_{2,2}^{(\mathrm{M })}= 1/2$, $\L_{2,3}^{(\mathrm{M })}=0.211...$, and $\L_{3,3}^{(\mathrm{M })}= 1/2$. The detailed computation is shown as follows.

\begin{table}
\begin{center}
\begin{tabular}{c | c | c | c | c}
\hline \hline
~$m$ ~& ~$\vphantom{\biggl\lbrace}~ \L_{m,2}^{(\mathrm{M})}$~ &~ $\L_{m,3}^{(\mathrm{M})}$ ~& ~$\L_{m,4}^{- (\mathrm{M})}$ ~ & ~ $\L_{m,4}^{+ (\mathrm{M})}$ \\ \hline \hline
$2$  & 1/2& 0.211..& 0  & 0 \\ \hline
$3$  &  1& 1/2& 1/4 & 1/2 \\ \hline
$4$ & & 1& 1/2  & 1/2 \\ \hline
 $5$ & & & 1  & 1 \\
\hline \hline
\end{tabular}
\end{center}
\caption{Lower bounds $\L_{m,d}^{(\mathrm{M})}$ in Eq. (\ref{eq:mubl}) are summarized for $m$ MUBs in $\mathbb{C}^d$, for $d=2,3,4$. }
\label{tab:MUBs_lowerbound}
\end{table}

In $d=2$, there is only one pair of MUBs, $\{B_1,B_2\}$, up to equivalence, which can be expressed as a pair of matrices,
\bea
\B_1 =  \left(
\begin{array}{cc}
1 & 0 \\
0 & 1
\end{array}
\right)~\mathrm{and}  ~\B_2 = \frac{1}{\sqrt{2}} \left(
\begin{array}{cc}
1 & 1  \\
1 & -1
\end{array}
\right),\nonumber
\eea
where the columns form the basis elements. Hence, for $m=2$, there is no need to optimize over all pairs of MUBs, and a numerical minimization is applied over all states $\ket{e}=\cos(\theta)\ket{0}+e^{i\phi}\sin(\theta)\ket{1}$ and $\ket{f}=\cos(\theta')\ket{0}+e^{i\phi'}\sin(\theta')\ket{1}$ in Eq. (\ref{eq:mubl}). This gives the bound $\L_{2,2}^{ (\mathrm{M })} = 1/2$, as shown in Table \ref{tab:MUBs_lowerbound}. The case $m=3$, i.e., a quantum $2$-design, for which  $\L_{3,2}^{ (\mathrm{M })} = 1 $ will be shown later using a connection to EWs.

In $d=3$, the complete set of four MUBs in matrix form are,
\bea
\B_1 & = & \left(
\begin{array}{ccc}
1 & 0 & 0\\
0 & 1 & 0\\
0 & 0 & 1
\end{array}
\right),  ~~
\B_2  =  \frac{1}{\sqrt{3}} \left(
\begin{array}{ccc}
1 & 1 & 1\\
1 & \omega & \omega^{2}\\
1 & \omega^{2} & \omega
\end{array}
\right), \nonumber\\
\B_3 & = & \frac{1}{\sqrt{3}} \left(
\begin{array}{ccc}
1 & 1 & 1\\
\omega & \omega^{2} & 1\\
\omega & 1 & \omega^{2}\end{array}
\right),~ ~
\B_4   =  \frac{1}{\sqrt{3}} \left(
\begin{array}{ccc}
1 & 1 & 1\\
\omega^{2} & 1 & \omega\\
\omega^{2} & \omega & 1
\end{array}
\right),\nonumber
\eea
where the columns form the basis elements. For each $m\leq 4$, there exists only one equivalence class of MUBs. That is, $\{\mathcal{B}_1,\mathcal{B}_2\}$ for $m=2$, and $\{\mathcal{B}_1,\mathcal{B}_2, \mathcal{B}_3\}$ when $m=3$. By performing a minimization of $I_{m,d}^{(\mathrm{M})}(\sigma_{\text{sep}})$ over all normalized states $\ket{e},\ket{f}\in\mathbb{C}^3$, we obtain the bounds given in Table \ref{tab:MUBs_lowerbound}.

For $d=4$, it is no longer true that there is a unique equivalence class of MUBs for each $m$, thus, we have in general,
\bea
\L_{m,4}^{-(\mathrm{ M})} \leq \L_{m,4}^{ + (\mathrm{ M})}.\nonumber
\eea
For pairs of MUBs, i.e., $m=2$, there exists a one-parameter family of equivalence classes, denoted by $\mathcal{P}(x)=\{\B_1, \B_2(x)\}$, and for triples of MUBs, i.e., $m=3$, there exists a three-parameter family of equivalence classes, namely,
\bea
\mathcal{T}(x,y,z)=\{\B_1,\B_2(x),\B_3(y,z)\}.\label{eq:mubtriple}
\eea
Here, the parameters take the values $x,y,z\in[0,\pi]$, and in matrix form, the bases can be expressed as
\bea
&& \B_1 = \left(
\begin{array}{cccc}
1 & 0 & 0 & 0 \\
0 & 1 & 0 & 0 \\
0 & 0 & 1 & 0 \\
0 & 0 & 0 & 1
\end{array}
\right),
\B_2(x)   =  \frac{1}{2} \left(
\begin{array}{cccc}
1 & 1 & 1 & 1 \\
1 & 1 & -1 & -1 \\
1 & - 1 & ie^{ix} & -ie^{ix} \\
1 & -1 & -ie^{ix} & ie^{ix}
\end{array}
\right),  \nonumber\\
&&  \mathrm{and}~~\B_3(y,z) = \frac{1}{2} \left(
\begin{array}{cccc}
1 & 1 & 1 & 1 \\
1 & 1 & -1 & -1 \\
-e^{iy} & e^{iy} & e^{iz} & -e^{iz} \\
e^{iy} & -e^{iy} & e^{iz} & -e^{iz}
\end{array}
\right),\label{eq:mub4}
\eea
 where the columns correspond to the basis vectors \cite{brierley10}. Then, for $x\neq x'$ the two sets $\mathcal{P}(x)$ and $\mathcal{P}(x')$ are inequivalent. Similarly, the two sets $\mathcal{T}(x,y,z)$ and $\mathcal{T}(x',y',z')$ for $(x,y,z)\neq (x', y', z')$ are inequivalent. When $m=2$, it turns out that, nevertheless, all pairs of MUBs provide the same lower bound, i.e., $\L_{2,4}^{\pm (\mathrm{M })} := \L_{2,4}^{(\mathrm{M })}$. However, since $\L_{2,4}^{(\mathrm{M})} =0$, no entangled state can be detected via this bound.

Next, for $m=3$, the lower bound varies according to our choice of triple $\mathcal{T}(x,y,z)$. Considering all possible sets of $3$ MUBs, we find
\bea
\L_{3,4}^{+(\mathrm{M })} & = & \max_{x,y,z}   \L_{3,4}^{(\mathrm{M})} (\mathcal{T}(x,y,z)) = \frac{1}{2},~\mathrm{and} \nonumber\\
\L_{3,4}^{-(\mathrm{M })}  & = & \min_{x,y,z}    \L_{3,4}^{(\mathrm{M})} (\mathcal{T}(x,y,z)) = \frac{1}{4}. \nonumber
\eea
These bounds are achieved for the triples $\mathcal{T}(\pi/2,0,0)$ and $\mathcal{T}(\pi/2,\pi/2,\pi/2)$, respectively. The only triple which extends to a larger set of MUBs is $\mathcal{T}(\pi/2,\pi/2,\pi/2)$. All other members of the three-parameter family are examples of unextendible MUBs. Hence, the unextendible MUBs detect more entanglement than the extendible triple since they provide tighter lower bounds.

There is only one equivalence class of MUBs for each $m=4,5$, given by $\mathcal{T}(\pi/2,\pi/2,\pi/2)\cup \{B_4\}$ and $\mathcal{T}(\pi/2,\pi/2,\pi/2)\cup \{B_4,B_5\}$, respectively, where,
\bea
\B_4 & = & \frac{1}{2} \left(
\begin{array}{cccc}
1 & 1 & 1 & 1 \\
i & -i & i & -i \\
-1 & - 1 & 1 & 1 \\
i & -i & -i & i
\end{array}
\right),~\mathrm{and} \nonumber\\
\B_5 & = & \frac{1}{2} \left(
\begin{array}{cccc}
1 & 1 & 1 & 1 \\
i & -i & i & -i \\
i & -i & -i & i \\
-1 & -1 & 1 & 1
\end{array}
\right).\label{eq:MUBsD=4}
\eea
Thus, since it is not necessary to optimize over collections of MUBs, we perform a minimization over product states to find $\L_{4,4}^{(\mathrm{M})} = 1/2$ and $\L_{5,4}^{(\mathrm{M})} = 1$. These bounds, including the case $m=4$, are summarized in Table \ref{tab:MUBs_lowerbound}.

\subsection{Lower and upper bounds on $I_{\widetilde{m},d }^{(\mathrm{S})}$ }
\label{subsec:sic}
We now consider measurements using SIC states to construct similar inequalities for $I_{\m,d}^{(\mathrm{S})}$ defined in Eq. (\ref{eq:sicc}). It is important to specify which SIC-POVM $S_{d^2}$ we use for our measurements, as for a given dimension $d$ they are usually not unique. Hence, the bounds we derive will depend explicitly on the given SIC-POVM.

 For a subset of $\m$ SICs, $S_{\m} = \{ |s_j\rangle \}_{j=1}^{\m}\subseteq S_{d^2}$ with $\m\leq d^2$, let $\U_{\m,d}^{(\mathrm{S})}$ and $\L_{\m,d}^{(\mathrm{S})}$ denote the upper and lower bounds,
\bea
\L_{\m,d}^{(\mathrm{S})} (S_{\m}) & = & \min_{\sigma_{\mathrm{sep}}} ~ I_{m,d}^{(\mathrm{S})} (\sigma_{\mathrm{sep}} :S_{\m} )\,, \label{eq:siclAppendix} \\
\U_{\m,d}^{(\mathrm{S})} (S_{\m})  & = & \max_{\sigma_{\mathrm{sep}}} ~ I_{m,d}^{(\mathrm{S})} (\sigma_{\mathrm{sep}} :S_{\m} )\,. \label{eq:sicuAppendix}
\eea
Since the lower and upper bounds may depend on which subset of $\m$ states are taken from the SIC-POVM $S_{d^2}$, let us introduce maximal and minimal bounds optimized over $\m$ collections of SICs, as follows,
\bea
\L_{\m,d}^{- (\mathrm{S})} & = & \min_{S_{\m} \subseteq S_{d^2}} \L_{\m,d}^{(\mathrm{S})} (S_{\m})\,,    \label{eq:sicl-} \\
\L_{\m,d}^{+  (\mathrm{S})} & = & \max_{S_{\m} \subseteq S_{d^2}} \L_{\m,d}^{(\mathrm{S})} (S_{\m})\,,    \ \label{eq:sicl+} \\
\U_{\m,d}^{- (\mathrm{S})} & = & \min_{S_{\m} \subseteq S_{d^2}} \U_{\m,d}^{(\mathrm{S})} (S_{\m}) \,,   \label{eq:sicu-}\\
\U_{\m,d}^{+ (\mathrm{S})} & = & \max_{S_{\m} \subseteq S_{d^2}} \U_{\m,d}^{(\mathrm{S})} (S_{\m})\,,    \label{eq:sicu+}
\eea
where $S_{d^2}$ is a given SIC-POVM in dimension $d$. Note that these optimizations can only be applied when the explicit form of the SIC-POVM is known, which is not the case in large dimensions. As shown below, it turns out that $I_{\m,d }^{(\mathrm{S})}$ satisfies the inequalities,
\bea
\L_{\m,d }^{ - (\mathrm{S })} \leq \L_{\m,d }^{ + (\mathrm{S })} ~ \leq ~ I_{\m,d }^{(\mathrm{S})} (\sigma_{\mathrm{sep}}) ~ \leq ~\U_{\m,d }^{ - (\mathrm{S })} \leq \U_{\m,d }^{ + (\mathrm{S })},\nonumber\\  \label{eq:siclu}
\eea
for all separable states $\sigma_{\mathrm{sep}}$. Entangled states are detected by violations of these inequalities. The tighter bounds only apply for a specific subset of SICs, i.e., the set $S_{\m}\subseteq S_{d^2}$ used to find $\L_{\m,d}^{+  (\mathrm{S})}$ or $\U_{\m,d}^{- (\mathrm{S})}$ in Eqs. (\ref{eq:sicl+}) and (\ref{eq:sicu-}). The weaker bounds apply for any subset of $\m$ states chosen from a particular SIC-POVM.

For the optimizations in Eqs. (\ref{eq:siclAppendix}) and (\ref{eq:sicuAppendix}) over separable states, it suffices to consider only product states due to the convexity of the set of separable states. Hence, given a set $S_{\m}$ of $\m$ SICs,
\bea
\L_{\m,d}^{(\mathrm{S})}  (S_{\m}) & = & \min_{|e\rangle, |f\rangle}  \sum_{|s_j\rangle \in S_{\m}}  |\langle s_j | e\rangle |^2 |\langle s_j | f\rangle |^2\,, \label{eq:sicle}\\
\U_{\m,d}^{(\mathrm{S})}  (S_{\m})  & = & \max_{|e\rangle} \sum_{|s_j\rangle \in S_{\m}} |\langle s_j | e \rangle|^4\,,\label{eq:sicue}
\eea
where $\ket{e},\ket{f}\in\mathbb{C}^d$. We have not yet found a systematic method to find these minimal and maximal bounds in general. In the following we derive the bounds for $d=2,3$ and optimize over all subsets of $\m$ SICs from a given SIC-POVM. However, for $d=4$, we only find suboptimal bounds.

\begin{table}
\begin{center}
\begin{tabular}{c | c | c}
\hline \hline
 $\widetilde{m}$ SICs~ & $\vphantom{\biggl\lbrace}~ \L_{\m,2}^{(\mathrm{S  })} $  & $  \U_{\m,2}^{(\mathrm{S })}$  \\ \hline \hline
2   ~&~ 0  ~&~   $  (\sqrt{3}+1)^2 /6$  \\ \hline
3 ~&~  4/15 ~&~ 4/3 \\ \hline
4 ~&~ 2/3 ~&~ 4/3 \\
\hline \hline
\end{tabular}
\end{center}
\caption{Lower and upper bounds $\L_{\m,2}^{ (\mathrm{S })}$ and $\U_{\m,2}^{ (\mathrm{S })}$ are shown for $\widetilde{m}=2,3,4$ in dimensions $d=2$. When $\m$ SICs are chosen from a set of $d^2$ states, there are $4 ! ( \m! \ (4-\m)!)^{-1} $ possible subsets of $\m$ SIC states. It turns out that for $d=2$, these bounds do not depend on the selection of $\m$ states.}
\label{tab:SICs_boundsd=2}
\end{table}

\subsubsection{Upper bounds $\U_{\m,d}^{(\mathrm{S})}$}

As previously mentioned, the bounds we derive will depend explicitly on the given SIC-POVM.  Here, we will only consider Heisenberg-Weyl SICs which are constructed from the Heisenberg-Weyl group, generated by the phase and cyclic shift operators (modulo $d$), which are defined as
\bea
Z\ket{j}=\omega^j\ket{j},\quad X\ket{j}=\ket{j+1}, \label{eq:HWoperators}
\eea
where $\omega=e^{2\pi i/d}$ and $\{\ket{j}\}_{j=0}^{d-1}$ is the standard basis of $\mathbb{C}^d$. A Heisenberg-Weyl SIC can then be constructed by taking the orbit of a fiducial vector $\ket{\psi_f}$, i.e.,
\bea
\ket{s_{a,b}}=e^{-iab\pi/d}X^aZ^b\ket{\psi_f},\label{eq:hwsic}
\eea
for $a,b=0,\ldots,d-1$.

For dimension $d=2$, there is a unique SIC-POVM, found in \cite{ref:zauner,ref:renes}, which can be generated from the fiducial vector
\bea
\ket{\psi_f}=\frac{1}{\sqrt{6}}\left(\sqrt{3+\sqrt{3}}\ket{0}+e^{\pi i/4}\sqrt{3-\sqrt{3}}\ket{1}\right).\nonumber
\eea
This SIC-POVM can be written more simply as the four vectors
\bea
|s_1\rangle & = & |0\rangle,  \nonumber \\
|s_2\rangle & = & \frac{1}{\sqrt{3}} (| 0\rangle + \sqrt{2} |1\rangle),  \nonumber \\
|s_3\rangle & = & \frac{1}{\sqrt{3}} ( e^{-\frac{i\pi}{3}} | 0\rangle + \sqrt{2} e^{\frac{i\pi}{3}} |1\rangle),  \nonumber \\
|s_4\rangle & = & \frac{1}{\sqrt{3}} ( e^{ \frac{i\pi}{3}} | 0\rangle + \sqrt{2} e^{ -\frac{i\pi}{3}} |1\rangle),\label{eq:sic4d2}
\eea
where $\{\ket{0},\ket{1}\}$ is the standard basis of $\mathbb{C}^2$. The states, which form a quantum 2-design, also form a tetrahedron in the Bloch sphere. It turns out that the upper bounds we calculate do not depend on which choice of $\m$ SICs from Eq. (\ref{eq:sic4d2}) we take. In particular, for both $\m=2,3$, the two bounds $\U_{\m,d}^{- (\mathrm{S })}$ and $\U_{\m,d}^{+ (\mathrm{S })}$ coincide, i.e., $\U_{\m,d}^{- (\mathrm{S })}=\U_{\m,d}^{+ (\mathrm{S })}$.
For  $\m=2$, the upper bound for any pair of SICs is
\begin{equation}
\U_{2,2}^{- (\mathrm{S })} = \U_{2 ,2}^{+ (\mathrm{S })}=\frac{1}{6}(\sqrt{3}+1)^2\,. \nonumber
\end{equation}
For $\m=3$, any subset of three SICs taken from Eq. (\ref{eq:sic4d2}) gives
\begin{equation}
\U_{3,2}^{-(\mathrm{S })}=\U_{3,2}^{+(\mathrm{S })}=\frac{4}{3}. \nonumber
\end{equation}
We note that it is also possible to find the vector $\ket{e}$ in Eq. (\ref{eq:sicue}) which attains these bounds. For example, given the set $\{\ket{s_1},\ket{s_3}\}$, then $\ket{e_{\text{max}}}=\kappa(\ket{s_1}+e^{\pi i/3}\ket{s_3})$, where $\kappa$ is a normalization factor. In fact, in all of the dimensions we investigate, given a set of $\m$ SICS, $\{\ket{s_j}\}$, the vector achieving the maximum takes the form $\ket{e_{\text{max}}}=\kappa(\sum_{j}e^{\pi i \lambda_j}\ket{s_j})$, where the summation is taken over all SICs from the set $\{\ket{s_j}\}$.

Now we move to dimension $d=3$, and choose the Hesse SIC \cite{ref:renes,ref:zauner}, which is generated by the Heisenberg-Weyl group from the fiducial vector
\begin{equation}
\ket{\psi_f}=\frac{1}{\sqrt{2}}\left(\ket{1}-\ket{2}\right). \nonumber
\end{equation}
Written explicitly, the nine SIC vectors are
\begin{align}
\ket{s_1}&=\frac{1}{\sqrt{2}}(\ket{1} - \ket{2}), \nonumber \\
\ket{s_2}&=\frac{1}{\sqrt{2}}(-\ket{0} + \ket{2}),  \nonumber  \\
\ket{s_3}&=\frac{1}{\sqrt{2}}(\ket{0} - \ket{1}), \nonumber  \\
\ket{s_4}&=\frac{1}{\sqrt{2}}(\omega\ket{1}-\omega^2\ket{2}), \nonumber  \\
\ket{s_5}&=\frac{1}{\sqrt{2}}( -\omega\ket{0} + \ket{2}), \nonumber  \\
\ket{s_6}&=\frac{1}{\sqrt{2}}( \omega^2\ket{0}-\ket{1}),\nonumber  \\
\ket{s_7}&=\frac{1}{\sqrt{2}}( \omega^2\ket{1} - \omega\ket{2}), \nonumber \\
\ket{s_8}&=\frac{1}{\sqrt{2}}( -\omega^2\ket{0} + \ket{2}), \nonumber \\
\ket{s_9}&=\frac{1}{\sqrt{2}}( \omega\ket{0} - \ket{1}), \label{eq:sic9d3}
\end{align}
where $\omega=\exp(2 \pi i/3)$ and $\{\ket{0},\ket{1},\ket{2}\}$ is the standard basis of $\mathbb{C}^3$.

\begin{table}
\begin{center}
\begin{tabular}{c | c | c | c | c}
\hline \hline
 $\m $ SICs ~&~ $ \vphantom{\biggl\lbrace}  \L_{\m,3}^{ - (\mathrm{S } ) } $ ~ &~ $\L_{\m, 3}^{+(\mathrm{S })} $  ~& ~ $ \U_{\m, 3}^{+ ( \mathrm{S } ) } $ ~ & ~  $\U_{\m, 3}^{- (\mathrm{S })}$ \\ \hline \hline
3  & 0      & 0 &1.25414... & 9/8  \\ \hline
4   &0      & 0 &1.39952... &  1.25414... \\ \hline
5  & 0      & 0 &1.46301... &  1.39952... \\ \hline
6  & 0      & 0.1123 & 3/2   &  1.48175... \\ \hline
7  & 3/20 & 3/20  & 3/2      &   3/2 \\ \hline
8  & 3/8   & 3/8    & 3/2      &    3/2 \\ \hline
9  & 3/4   &  3/4   & 3/2      &    3/2 \\
\hline \hline
\end{tabular}
\end{center}
\caption{In dimension $d=3$, we use the Hesse SIC-POVM defined by the 9 SIC vectors of Eq. (\ref{eq:sic9d3}). Both the lower and upper bounds depend on the choice of $\m$ SIC states, although when $m$ is large the upper bounds become independent of the choice and number of SICs. The minimal and maximal bounds of $\L_{\m,d}^{\pm (\mathrm{S})}$ and $\U_{\m,d}^{\pm (\mathrm{S})}$ are shown for $\m=3,4,5,6,7,8,9$, some of which are obtained numerically.}
\label{tab:SICs_boundsd=3}
\end{table}

In contrast to $d=2$, here we find that the two bounds $\U_{\m,d}^{- (\mathrm{S })}$ and $\U_{\m,d}^{+ (\mathrm{S })}$ do not in general coincide. We calculate the bound $\U_{\m,d}^{(\mathrm{S})} (S_{\m})$ in Eq. (\ref{eq:sicuAppendix}) for all subsets of $\m$ SICs from the $9$ vectors in Eq. (\ref{eq:sic9d3}), and find that for each $\m$ there are at most two different bounds. Thus, the smallest of the two bounds must coincide with $\U_{\m,d}^{- (\mathrm{S})}$, while the largest must coincide with  $\U_{\m,d}^{+ (\mathrm{S})}$.

When $\m=3$, we find two upper bounds for $\U_{\m,d}^{(\mathrm{S})} (S_{\m})$ depending on our choice of three SICs, $\{\ket{s_i},\ket{s_j},\ket{s_k}\}$, from Eq. (\ref{eq:sic9d3}). Denoting this set by the indices $(i,j,k)$, we find that  $(1,2,3)$, $(1,4,7)$, $(1,5,9)$, $(1,6,8)$, $(2,4,9)$, $(2,5,8)$, $(2,6,7)$, $(4,5,6)$ and $(7,8,9)$, give an upper bound of $9/8$. The remaining sets give a numerical upper bound of $1.25414$. Thus, the values of $\U_{\widetilde{3},3}^{+ (\mathrm{S })}$ and $\U_{\widetilde{3},3}^{+ (\mathrm{S })}$ are given by
\bea
\U_{\widetilde{3},3}^{+ (\mathrm{S })}=1.25414\ldots, \label{eq:bound1m3}
\eea
and
\bea
\U_{\widetilde{3},3}^{ - (\mathrm{S })}=9/8.\label{eq:bound2m3}
\eea

For $\m=4$, we also find that the minimal and maximal bounds do not coincide, although all subset of 4 SICs give one of two possible bounds. For example, the set $\{\ket{s_1},\ket{s_2},\ket{s_3},\ket{s_4}\}$ gives the bound
\bea
\U_{ 4,3}^{- (\mathrm{S })}=1.25414\ldots,\label{eq:bound1m4}
\eea
while from the set $\{\ket{s_1},\ket{s_2},\ket{s_4},\ket{s_5}\}$ we have,
\bea
\U_{4,3}^{+ (\mathrm{S })}=1.39952\ldots\label{eq:bound2m4}
\eea

When $\m=5$, two upper bounds for Eq. (\ref{eq:sicuAppendix}) also exist for all combinations of five SICs. For example, $\{\ket{s_i}\}_{i=1}^{5}$ gives the bound
\bea
\U_{5,3}^{+ (\mathrm{S })}=1.46301\ldots,\label{eq:bound1m5}
\eea
while for the set $\{\ket{s_1},\ket{s_2},\ket{s_3},\ket{s_4},\ket{s_7}\}$ we have,
\bea
\U_{ 5,3 }^{- (\mathrm{S })}=1.39952...\label{eq:bound2m5}
\eea

For $\m=6$, again there are two distinct upper bounds for all combination of six SICs. For example, $\{\ket{s_i}\}_{i=1}^{6}$ gives
\bea
\U_{6,3}^{+ (\mathrm{S })}=\frac{3}{2}\,,\label{eq:bound1m6}
\eea
and for $\{\ket{s_1},\ket{s_2},\ket{s_3},\ket{s_4},\ket{s_5},\ket{s_7}\}$ we have
\bea
\U_{ 6,3 }^{- (\mathrm{S })}=1.48175\ldots\label{eq:bound2m6}
\eea

When $\m=7$, the two bounds coincide for all subsets of 7 SICs such that
\bea
\U_{ 7,3 }^{- (\mathrm{S})}=\U_{7,3 }^{+ (\mathrm{S })}=\frac{3}{2}.\label{eq:bound1m7}
\eea

Finally, for $\m=8$, any set of 8 SIC vectors yields the same bound, hence,
\begin{equation}
\U_{ 8,3 }^{-  (\mathrm{S})}=\U_{8,3}^{+ (\mathrm{S})}=\frac{3}{2}.\label{eq:bound1m8}
\end{equation}

Next, for dimension $d=4$, we choose the SIC-POVM generated from the fiducial vector
\begin{equation}
\ket{\psi_f}=\frac{1}{2\sqrt{3+\Gamma}}(\alpha_+\ket{0}+\beta_+\ket{1}+\alpha_-\ket{2}+\beta_-\ket{3})\,,\label{eq:sicfd4}
\end{equation}
where $\alpha_{\pm}=1\pm e^{-i\pi/4}$, $\beta_{\pm}=e^{i\pi/4}\pm i\Gamma^{-3/2}$, and $\Gamma=(\sqrt{5}-1)/2$ is the golden ratio \cite{ref:renes,ref:zauner}.
We have tested several possible subsets $S_{\m}$ for each $\widetilde{m}$ and found that the bound $\U_{\widetilde{m},d}^{(\mathrm{S})}(S_{\m})$ in Eq. (\ref{eq:siclu}) takes a number of different values depending on the set $S_{\m}$. Let $\ket{s_{a,b}}\in\mathbb{C}^4$ be the SIC states defined in Eq. (\ref{eq:hwsic}) generated by the fiducial vector of Eq. (\ref{eq:sicfd4}). Choosing the set $S_{\m}$ as the sets $\{\ket{s_{0,0}},\ket{s_{0,1}},\ket{s_{0,2}}\}$, $\{\ket{s_{0,0}},\ket{s_{0,1}},\ket{s_{0,2}},\ket{s_{0,3}}\}$, $\{\ket{s_{0,0}},\ket{s_{0,1}},\ket{s_{0,2}},\ket{s_{0,3}},\ket{s_{1,0}}\}$, etc., for $\m=3,4,5,\ldots$, we present the suboptimal bounds $\U_{\m,d}^{(\mathrm{S})}(S_{\m})$ in Table \ref{tab:SICs_boundsd=4}.

\begin{table}
\begin{center}
\begin{tabular}{c | c | c}
\hline \hline
 $\m $ SICs~ &~ $\vphantom{\biggl\lbrace} \L_{\m,4}^{(\mathrm{S })}(S_{\m}) $ ~&~ $ \U_{\m, 4 }^{(\mathrm{S })}(S_{\m})$\\ \hline \hline
 3  & 0 &  1.1476 \\ \hline
4   & 0 & 1.2676 \\ \hline
5&  0 & 1.3766 \\ \hline
6 & 0 & 1.4521 \\ \hline
 7  & 0.0067 & 1.4723 \\ \hline
8& 0.0279 & 1.4902 \\ \hline
9& 0.0325 & 1.5556 \\ \hline
10& 0.0693 & 1.5763  \\ \hline
11& 0.0719 & 1.5881\\ \hline
12& 0.1436 & 1.5935 \\ \hline
13& 0.2031 & 1.6 \\ \hline
14& 0.2285 & 1.6  \\ \hline
15& 0.4363 & 1.6 \\ \hline
16& 4/5 & 1.6  \\
\hline \hline
\end{tabular}
\end{center}
\caption{For $\m$ SIC states in $\mathbb{C}^4$, suboptimal lower and upper bounds, denoted by $\L_{\m,4}^{(\mathrm{S})}(S_{\m})$ and $\U_{\m,4}^{(\mathrm{S})}(S_{\m})$, are presented, for the quantity $I_{\m,4}^{(\mathrm{S})}$. These bounds are satisfied for all separable states, provided a \emph{specific} subset $S_{\m}$ of $\m$ SICs is chosen from the SIC-POVM generated by the fiducial vector of Eq. (\ref{eq:sicfd4}). The subsets we apply are given below Eq. (\ref{eq:sicfd4}). 
}
\label{tab:SICs_boundsd=4}
\end{table}

\subsubsection{Lower bounds $\L_{\m,d}^{(\mathrm{S})}$}

We now apply similar techniques to calculate the lower bound of Eq. (\ref{eq:sicle}) for $I_{\m,d}^{(\mathrm{S})}$. We will use the same SIC-POVMs as defined in the previous section.
For dimensions $d=2$ we find that $\L_{\widetilde{m},2}^{+(\mathrm{S })} =\L_{\widetilde{m}, 2}^{-(\mathrm{S })}$ for all $\widetilde{m}=2,3,4$, hence the bounds in Table \ref{tab:SICs_boundsd=2} are both optimal and apply for any choice of SICs.
\\

For $d=3$ and $\widetilde{m}=0,\ldots,5$ we find that $\L_{\widetilde{m},d}^{-(\mathrm{S})}=\L_{\widetilde{m},d}^{+(\mathrm{SIC})}=0$. When $\widetilde{m}=6$, we have
\bea
\L_{6,3}^{-(\mathrm{S})} =0,
\eea
and
\bea
\L_{6 ,3}^{+(\mathrm{S})} =0.1123.
\eea
In fact, for any choice of $6$ vectors from the nine SICs in Eqs. (\ref{eq:sic9d3}), the lower bound is either 0 or 0.1123. For example, given the set $S_{\m}=\{\ket{s_i}\}_{i=1}^6$, we obtain  $\L_{6,3}^{(\mathrm{S})}(S_{\m}) =0$, while for $S_{\m}=\{\ket{s_1},\ket{s_2},\ket{s_3},\ket{s_4},\ket{s_5},\ket{s_7}\}$ we have $\L_{6 ,3}^{(\mathrm{S})}(S_{\m}) =0.1123$.

For $\widetilde{m}=7$, the bounds are independent of the choice of SICs and attain the value $\L_{\widetilde{m},d}^{-(\mathrm{S})}=\L_{\widetilde{m},d}^{+(\mathrm{S})}=0.15$. Finally, for $\widetilde{m}=8$ the bounds take the value $\L_{\widetilde{m},d}^{\pm(\mathrm{S})}  =0.375$.

In dimension $d=4$, suboptimal lower bounds are presented in Table \ref{tab:SICs_boundsd=4} using the same sets of SICs as the upper bounds, i.e., $\{\ket{s_{0,0}},\ket{s_{0,1}},\ket{s_{0,2}}\}$, $\{\ket{s_{0,0}},\ket{s_{0,1}},\ket{s_{0,2}},\ket{s_{0,3}}$ etc., for $\m=3,4,\ldots$

\section{On the capability of detecting entangled states}

To summarize, we have derived inequalities given in Eqs. (\ref{eq:mublu}) and (\ref{eq:siclu}), for sets of MUBs and SICs, respectively.
First, for \emph{any} set of $m$ MUBs in dimension $d$, the quantity $I_{m,d}^{(\mathrm{M })}$ is bounded above and below, for all separable states, by
\bea
\L_{m,d}^{-(\mathrm{M })}\leq I_{m,d}^{(\mathrm{M })}(\sigma_{\text{sep}})\leq 1+\frac{m-1}{d},\label{eq:ulboundmubs}
\eea
where the values $\L_{m,d}^{-(\mathrm{M })}$ are given in Table \ref{tab:MUBs_lowerbound} for $d=2,3,4$. Note that $\L_{m,d}^{-(\mathrm{M })}=\L_{m,d}^{(\mathrm{M })}$ for $d=2,3$, due to the existence of only one equivalence class of $m$ MUBs. We also provide a tighter lower bound $\L_{m,d}^{+(\mathrm{M })}$ when $d=4$, which only holds true if we restrict the choice of MUBs to a specific set. When $m=3$, this triple of MUBs is given by $\mathcal{T}(\pi/2,\pi/2,\pi/2)\in\mathcal{T}(x,y,z)$, as defined in Eq. (\ref{eq:mubtriple}). From an experimental perspective, this triple of MUBs is more useful for entanglement detection since it detects a larger set of entangled states than any other member of the family $\mathcal{T}(x,y,z)$.

The situation is more complicated for SICs. First we must specify which SIC-POVM we apply in dimension $d$. We then show in dimensions $d=2,3$, that for \emph{any} subset of $\m$ SICs, the quantity $I_{\m,d}^{(\mathrm{S })}$ is bounded above and below, for all separable states, by
\bea
\L_{\m,d}^{-(\mathrm{S })}\leq I_{\m,d}^{(\mathrm{S })}(\sigma_{\text{sep}})\leq \U_{\m,d}^{+(\mathrm{S })},
\eea
where the values $\L_{\m,d}^{-(\mathrm{S })}$ and $\U_{\m,d}^{+(\mathrm{S })}$ are given in Tables \ref{tab:SICs_boundsd=2} and \ref{tab:SICs_boundsd=3}, for dimensions $d=2,3$, respectively. Note that in $d=2$, $\L_{\m,d}^{-(\mathrm{S })}=\L_{\m,d}^{(\mathrm{S })}$ and $\U_{\m,d}^{+(\mathrm{S })}=\U_{\m,d}^{(\mathrm{S })}$. For $d=3$ we can derive tighter upper and lower bounds,
\bea
\L_{\m,3}^{+(\mathrm{S })}\leq I_{\m,3}^{(\mathrm{S })}(\sigma_{\text{sep}})\leq \U_{\m,3}^{-(\mathrm{S })},\label{eq:ulboundssic1}
\eea
as sumarized in Table \ref{tab:SICs_boundsd=3}. However, Eq. (\ref{eq:ulboundssic1}) only applies for a specific set of $\m$ SICs, which we have specified explicitly in the derivations above. For $d=4$ we are unable to find upper and lower bounds which apply for \emph{any} subset of $\m$ SICs, however, we do find suboptimal bounds which apply for a specified set of $\m$ SICs, namely
\bea
\L_{\m,d}^{0(\mathrm{S })}\leq I_{\m,d}^{(\mathrm{S })}(\sigma_{\text{sep}})\leq \U_{\m,d}^{0(\mathrm{S })},\label{eq:ulboundssic2}
\eea
where the bounds are given in Table \ref{tab:SICs_boundsd=4}. To apply these bounds experimentally, it is required that the measurements correspond to the specified set of SICs.

We will also prove later that for a complete set of $(d+1)$ MUBs and $d^2$ SICs, which correspond to quantum 2-designs, the bounds simplify to
\bea
1 ~ \leq ~ I_{d+1,d}^{(\mathrm{M})}(\sigma_{\mathrm{sep}}) ~ \leq ~  2\,,  \label{eq:mubinqeq1}
\eea
and
\bea
\frac{d}{d+1} ~ \leq ~ I_{d^2,d}^{(\mathrm{S})}(\sigma_{\mathrm{sep}}) ~ \leq ~  \frac{2d}{d+1}.  \label{eq:sicineq1}
\eea

We now show that the inequalities above, for $I_{m,d}^{(\mathrm{M })}$ and $I_{\m,d}^{(\mathrm{S })}$, detect a larger set of entangled states as the number of measurements $m$ increases. In particular, we highlight the following result:\\

{\bf Remark 1}. As more MUBs and SICs are applied to the detection criterion, the stronger the capability of detecting entangled states.\\

For a graphical illustration of this phenomenon we refer the reader to Figures \ref{fig:picmub} and \ref{fig:picsic}.

\subsection{Examples: Symmetric States}

To demonstrate the observation made in Remark 1, we consider a particular class of bipartite $(d\times d)$-dimensional quantum states, the so-called symmetric states, and analyse their behaviour with respect to our detection criterion. The first set of states we investigate are the Werner states,
\bea
\rho_{\mathrm{W}} (p) & = & p \frac{2 }{ d(d+1)} \Pi_{\mathrm{sym}}  + (1-p) \frac{2  }{d(d-1)} \Pi_{\mathrm{asym}}\,,~~\label{eq:wernerstate}
\eea
where $p\in [0,1]$. Werner states are separable for $p\geq1/2$ and entangled if $p<1/2$.  We also consider the bipartite isotropic states which are invariant under $U\otimes U^{*}$,
\bea
\rho_{\mathrm{iso}}(q) & = & q  | \Phi^{+} \rangle \langle \Phi^{+} |  + (1-q) \frac{\mathbbm{1}\otimes\mathbbm{1}  }{d^2}\,, ~~\label{eq:isotropicstate}
\eea
where $q\in [0,1]$ and $| \Phi^{+}\rangle$ denotes a maximally entangled state. Isotropic states are entangled if and only if $q > 1/(d+1)$. Both of these symmetric states are non-positive under the partial transpose if and only if they are entangled. Once they are PPT, i.e. separable, Werner states can be converted to isotropic states, and vice versa, by the partial transpose.

In Ref. \cite{ref:spengler}, it is shown that an isotropic state is entangled if and only if it violates the upper bound in Eq. (\ref{eq:mubinqeq1}). For the Werner states defined in Eq. (\ref{eq:wernerstate}), it is straightforward to compute the following,
\bea
I_{m,d}^{(\mathrm{M})} (\rho_{\mathrm{W}}(p)) = \sum_{k=1}^{m}\sum_{i=1}^{d} \mathrm{Pr}(i,i|\B_k , \B_k)  = \frac{2pm}{ d+1 }, ~~~\label{eq:wbound}
\eea
where we note that $ \mathrm{Pr}(i,i|\B_k , \B_k)  = 2p ( d+1 )^{-1}$. For $m=d+1$, we have $I_{d+1,d}^{(\mathrm{M})} (\rho_{\mathrm{W}}(p)) = 2p$. From Eq. (\ref{eq:mubinqeq1}), it follows that the Werner states $\rho_{W}(p)$ violate the lower bound if $p<1/2$, and hence the criterion coincides with the exact separability conditions. We also remark that the upper bound from Eq. (\ref{eq:mubinqeq1}) has been used to detect bound entangled states \cite{ref:hiesmayr2,ref:hiesmayr1}.

In Fig. \ref{fig:picmub}, the inequalities of Eq. (\ref{eq:ulboundmubs}) for $I_{2,3}^{(\mathrm{M})}$, $I_{3,3}^{(\mathrm{M})}$, and $I_{4,3}^{(\mathrm{M})}$ are applied to detect entangled Werner and isotropic states in dimension $d=3$. It is shown that as $m$ decreases, i.e., as fewer MUBs are measured, the range of the entangled states detected becomes smaller.

For SICs, it is also straightforward to compute the following quantities,
\bea
I_{\m,d}^{(\mathrm{S })} ( \rho_{\mathrm{W}} (p) ) & = & \frac{ 2p \m }{ d(d+1) }\,, \label{isoboundsic} \\
I_{\m,d}^{(\mathrm{S  })} (\rho_{\mathrm{iso}} (q) ) & = & \frac{ \m}{d^2 } (q (d-1) + 1), \label{isoboundmub}
\eea
for Werner and isotropic states. We can then determine for which parameters the states satisfy the inequalities in Eq. (\ref{eq:sicineq1}). When $\m=d^2$, it follows that the upper bound in Eq. (\ref{eq:sicineq1}) is violated by all entangled isotropic states, and the lower bound by all entangled Werner states, that is, for $q > 1/(d+1)$ and $p< 1/2$, respectively. Thus, $I_{d^2,d}^{(\mathrm{S })}$ tightly characterizes entangled Werner and isotropic states.

\begin{figure}
\includegraphics[scale=0.24]{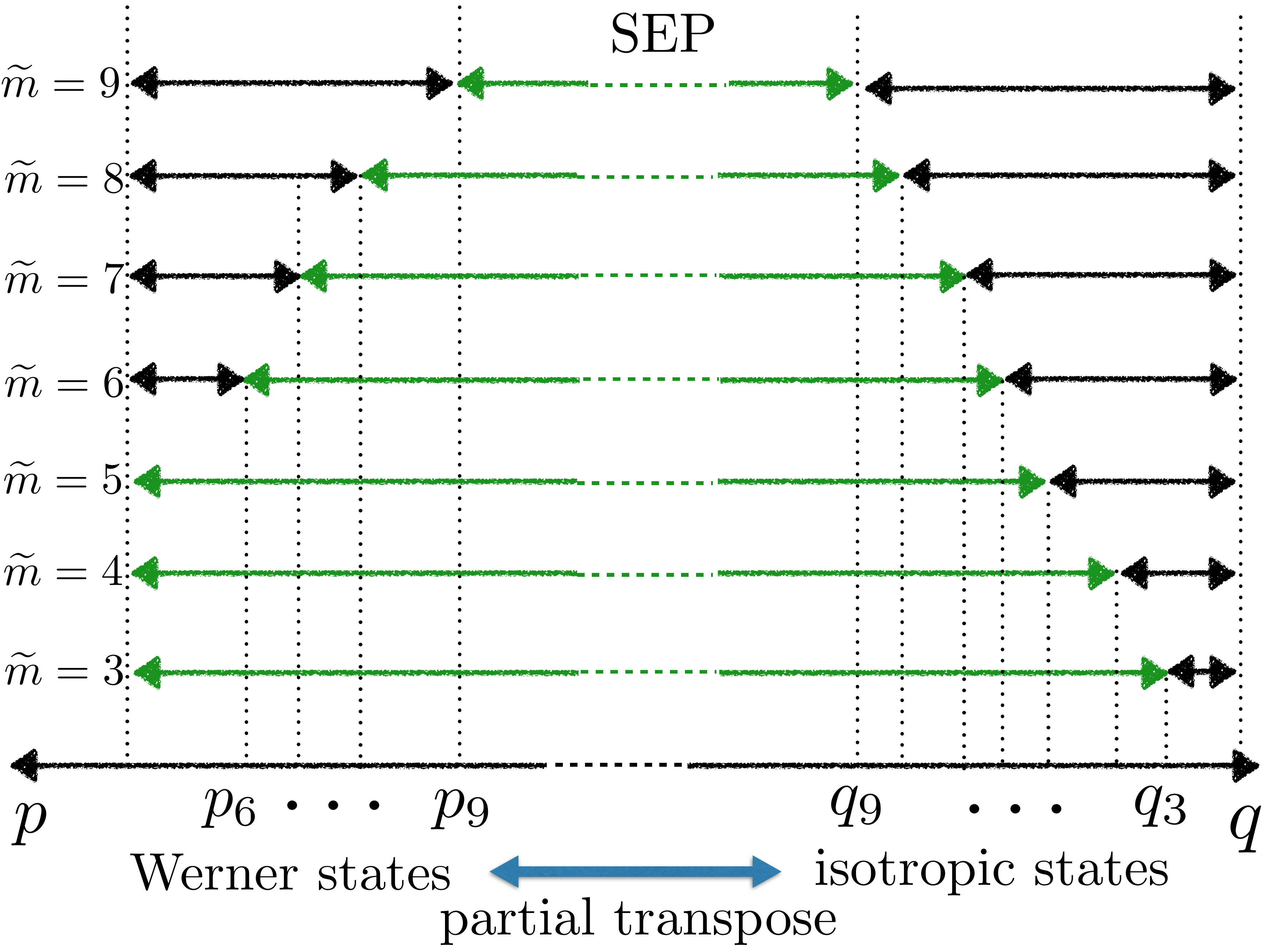}
\caption{ Inequalities for $I_{2,3}^{(\mathrm{S})}$,  $I_{3,3}^{(\mathrm{S})}$, and $I_{4,3}^{(\mathrm{S})}$ are applied to detect the entangled Werner and isotropic states defined in Eqs. (\ref{eq:wernerstate}) and (\ref{eq:isotropicstate}). The upper $(\U_{\m,d}^{-(\mathrm{S})})$ and lower $(\L_{\m,d}^{+(\mathrm{S})})$ bounds can be found in Table \ref{tab:SICs_boundsd=3}. Green colored lines show the range of states satisfying the inequalities, including all separable states. Lower bounds are violated by entangled Werner states and upper bounds by entangled isotropic states, with the critical values for the parameter $p$ and $q$ given as: $p_6 = 0.11$, $p_7 = 0.13$, $p_8 = 0.28$, $p_9 = 0.5$, $q_3 =  1.19$, $q_4 =  0.91$, $q_5 =  0.76$, $q_6 =  0.61$, $q_7 =  0.46$, $q_8 = 0.34$, and $q_9 =  0.25$.  }
\label{fig:picsic}
\end{figure}

The inequalities given in Eqs. (\ref{eq:ulboundssic1}) and (\ref{eq:ulboundssic2}) with $\m<d^2$ are applied to detect entangled states as follows. Given $\m$, and the values of $I_{\m,d}^{(\mathrm{S})}$ from Eqs. (\ref{isoboundsic}) and (\ref{isoboundmub}), a violation of either Eqs. (\ref{eq:ulboundssic1}) and (\ref{eq:ulboundssic2}) implies the states are entangled. Critical values of $p$ and $q$ that lead to violations of the inequality (\ref{eq:ulboundssic2}) for $d=3$ are shown in Fig. \ref{fig:picsic}, and are denoted by $\mathrm{L}_{\m}$ and $\mathrm{U}_{\m}$, respectively. One can naturally expect that a larger value of $\m$ implies a higher capability of detecting entangled states, as is indicated in the Figure.


\section{Relations: EWs and Quantum $2$-design}
\label{sec:ex}

In the following, we summarize the relationship between EWs and the structural physical approximation (SPA) to an EW. Both the EW and its SPA are equivalent in the sense that they detect the same set of entangled states. In particular, we show that when the SPA is applied to an EW constructed by the partial transpose, the resulting operator coincides with a quantum $2$-design. Furthermore, we derive the upper and lower bounds of $I_{m,d}^{(\mathrm{M})}$ and $I_{\m,d}^{(\mathrm{S})}$ for a full set of MUBs and SICs, respectively.


\subsection{EWs and their equivalent construction}

Let $\B(\H)$ denote the set of bounded operators in the Hilbert space $\H$. A Hermitian operator $W \in \B(\H \otimes \H)$ is an EW if it satisfies
\bea
&& \tr [W \sigma_{\mathrm{ sep }} ] \geq 0,~~\mathrm{for~all~separable~states}~\sigma_{\mathrm{sep}}\,, \nonumber \\
&& \tr [W \rho] < 0\,, ~~\mathrm{for~some~entangled ~states}~ \rho. \label{eq:ew}
\eea
EWs can be realized with positive-operator-valued-measures (POVMs). Given a witness $W$, one can find its decomposition
\bea
W = \sum_{i=1}^n c_i\; M_i\,, \label{eq:w}
\eea
for some $n>1$, with POVM elements $\{ M_i \}_{i=1}^n$. If this does not form a resolution of the identity operator, i.e., $\sum_{i=1}^{n} M_i < I$, then let $M_0$ denote the positive operator $M_0 = I  -\sum_{i=1}^n M_i$ such that one can construct a complete measurement $\{ M_i \}_{i=0}^{n}$.

Given an ensemble of identical quantum states $\rho$, on which individual measurements are performed, one obtains the probability distribution $\mathrm{Pr}(i | \rho ) = \tr[M_i \rho]$ for the set of detectors described by the POVM elements $M_i$. Collecting all outputs, one can compute
\bea
\tr[W\rho] = \sum_{i=1}^n c_i \mathrm{Pr} (i | \rho). \label{eq:formalism}
\eea
If Eq. (\ref{eq:formalism}) yields a negative value, we unambiguously conclude that the given state $\rho$ is entangled.

As it is mentioned above, EWs can be factorized into local observables, that is, local measurements.  A POVM element of local measurements can be written as,
\bea
M_i = M_{a}^x \otimes M_b^{y}\,, \nonumber
\eea
with indices $i= (x,y,a,b)$, where $M_{a}^{x}$ denotes a POVM element having outcome $a$ for measurement setting $x$. Suppose that witness $W$ has a decomposition containing only local measurements, i.e.,
\bea
W = \sum_{a,b,x,y} c_{a,b}^{x,y} M_{a}^{x}\otimes M_{b}^{y}.\nonumber
\eea
Then, the detection scheme with local measurements is given by the relation,
\bea
\tr[W \rho] & = & \sum_{a,b,x,y} c_{a,b}^{x,y}~ \mathrm{Pr}(a,b | x, y)\,,\label{eq:geinq}
\eea
where $\mathrm{Pr}(a,b | x, y) = \tr[ M_{a}^x \otimes M_b^{y} \rho]$ and the parameters $\{ c_{a,b}^{x,y} \}$ can be found from the witness $W$.

We now introduce an equivalent scheme for detecting entangled states by modifying an EW as follows. Let $X \in \B(\H \otimes \H)$ denote a non-negative, full-rank and unit-trace operator. Then, for a witness $W$, and an operator $X$, we define the following transformation,
\bea
W_X (p) = (1-p)\, W + p\, X \,, \label{eq:WX}
\eea
with parameter $0\leq p \leq 1$. Note that we have the relation, $W = W_{X} (p=0)$. Since $X$ is non-negative and of full-rank, it holds that for all separable states $\sigma_{\mathrm{sep}}$,
\bea
 \tr[W_X (p) \sigma_{\mathrm{sep}}]  & \geq & p~ m_s(X)\,, \label{eq:des}
\eea
where $m_s (X) = \min_{\sigma_{\mathrm{sep}}} \tr[X \sigma_{\mathrm{sep}}]$. In the minimization of
\bea
m_s (X) = \min_{|a\rangle,|b\rangle} \tr[|a\rangle \langle a|\otimes |b\rangle \langle b| X] \,,\label{eq:ms}
\eea
it sufficed to consider product states since mixing does not decrease the norm of the above quantity.

Inequalities satisfied by separable states can therefore be constructed from Eq. (\ref{eq:des}) as follows. Assume that $W_X (p)$ has a separable decomposition,
\bea
W_X(p) = \sum_{a,b,x,y} \widetilde{c}_{a,b}^{~x,y}\; M_{a}^x \otimes M_{b}^y\,, \label{eq:wxp}
\eea
for some fixed $p$, and let $P (a,b,|x,y) = \tr [M_{a}^x \otimes M_{b}^y \rho]$, for a given state $\rho$. Then it follows directly from Eq. (\ref{eq:des}) that the inequality
\bea
\sum_{x, y,a,b } \widetilde{c}_{a,b}^{~x,y}\; P(a,b | x,y ) \geq p\; m_s (X)\,,  \label{eq:inq}
\eea
is satisfied for all separable states. A violation of the inequality leads to the conclusion that the given quantum state $\rho$ is entangled.





\subsubsection{Lower bounds $\L_{d^2, d}^{(\mathrm{S })}$ and $\L_{d+1, d}^{(\mathrm{M })}$ }

We now derive the lower bounds $\L_{d^2, d}^{(\mathrm{S })}$ and $\L_{d+1, d}^{(\mathrm{M})}$ for sets of $d^2$ SIC vectors and $(d+1)$ MUBs, i.e., quantum $2$-designs. First, we consider Eq. (\ref{eq:WX}) with the following operators,
\bea
W = (\mathrm{id}\otimes T) [ | \Phi^{+ } \rangle\langle \Phi^{+} | ],~\mathrm{and}~X_0 = \frac{1}{d^2} \mathbbm{1}_d \otimes \mathbbm{1}_d\,,  \nonumber
\eea
where $|\Phi^{+}\rangle =  \sum_{i=1}^{d} | ii\rangle /\sqrt{d}$, is the maximally entangled state in $\mathbb{C}^d\otimes\mathbb{C}^d$. Note that $W = | \Phi^{+ } \rangle\langle \Phi^{+} |^{\Gamma}$, where $\Gamma$ denotes the partial transpose, corresponds to the permutation operator $\Pi$, i.e., $W=  d^{-1} (\Pi_{\mathrm{sym}}  - \Pi_\mathrm{asym} )$ where $\Pi_{\mathrm{sym}}$ and $\Pi_{\mathrm{asym}}$ denote the projectors onto the symmetric and antisymmetric subspaces. We also fix $p^{*} = d(d+1)^{-1}$ so that $W_{X_0} (p^*)$ is non-negative. This is called the structural physical approximation (SPA) of the witness \cite{intro1}.
Finally, Eq. (\ref{eq:WX}) can be expressed as,
\bea
W_{X_{0}} (p^*) = \frac{2}{d(d+1)} \Pi_{\mathrm{sym}}. \label{eq:WT}
\eea
Thus, it follows that the lower bound for separable states in Eq. (\ref{eq:des}) is given by,
\bea
\tr[W_{X_0} (p^{*}) \sigma_{\mathrm{sep}} ] \geq \min_{\sigma_{\mathrm{sep}}} p^{*} m_s (X_0) = \frac{1}{d(d+1)},
\label{eq:dlb}
\eea
where we have used the simple observation that $m_{s}(X_0) =   d^{-2}$. It is clear that the resulting witness in Eq.~(\ref{eq:WT}) corresponds to a quantum $2$-design (cf. Eq.~(\ref{eq:2design})), and can therefore be decomposed using a full set of $(d+1)$ MUBs or $d^2$ SICs. Thus, we can derive an inequalities of the form given in Eq. (\ref{eq:inq}) using MUBs and SICs.

First, if we consider a quantum $2$-design formed from a collection of $d^2$ SIC vectors, then Eq. (\ref{eq:WT}) can be decomposed as,
\bea
 \frac{2}{d(d+1)} \Pi_{\mathrm{sym}}  = \frac{1}{d^2} \sum_{j=1}^{d^2} | s_j\rangle \langle s_j | \otimes |s_j\rangle \langle s_j |\,.\nonumber
\eea
Since the left-hand-side of Eq. (\ref{eq:dlb}) can be decomposed in terms of SICs, the inequality can be rewritten as,
\bea
{I_{d^2,d}^{(\mathrm{S})}} =  \sum_{j=1}^{d^2} \mathrm{Pr}(j, j  | S_{d^2}, S_{d^2})  \geq  \frac{d}{d+1}  = \L_{d^2,d}^{(\mathrm{S})}. \nonumber
\eea
Thus, we have derived the lower bound $\L_{d^2,d}^{(\mathrm{S})} = d(d+1)^{-1}$.

Next, let us consider the case when the quantum $2$-design in Eq. (\ref{eq:WT}) is decomposed using a set of $(d+1)$ MUBs, i.e.,
\bea
 \frac{2}{d(d+1)} \Pi_{\mathrm{sym}}  = \frac{1}{d(d+1)} \sum_{k=1}^{d+1} \sum_{i=1}^{d} | b_{i}^{k} \rangle \langle  b_{i}^{k} | \otimes | b_{i}^{k} \rangle \langle   b_{i}^{k}  |\,.\nonumber
\eea
The left-hand-side of Eq. (\ref{eq:dlb}) is then written in terms of a set of MUBs, so that
\bea
{I_{d+1,d}^{(\mathrm{M})}} =  \sum_{k=1}^{ d+1 }\sum_{i=1}^d \mathrm{Pr}( i,i  | \B_k , \B_k)  \geq  1  = \L_{d+1,d}^{(\mathrm{M})}. \nonumber
\eea
Thus, we have shown that $\L_{d+1,d}^{(\mathrm{M})} =1$.

\subsubsection{ Upper bounds $\U_{d^2, d}^{(\mathrm{S })}$ and $\U_{d+1, d}^{(\mathrm{M })}$ }

The upper bounds $\U_{d+1, d}^{(\mathrm{M })}$ and $\U_{d^2, d}^{(\mathrm{S })}$ for a complete set of MUBs and SICs, respectively, can be derived via applications of the geometric mean \cite{ref:spengler,ref:li}. For the upper bound in Eq. (\ref{eq:sicuAppendix}), a separable state can be decomposed into a convex combination of product states, and due to the convexity it suffices to consider product states in the optimization. Therefore, we have
\bea
I_{d^2,d}^{(\mathrm{S })} (\sigma_{\mathrm{sep}}) & \leq & \max_{ |e\rangle,|f\rangle} \sum_{j=1}^{d^2} \tr[|s_j\rangle \langle s_j|^{\otimes 2}   ~| e \rangle \langle e | \otimes | f  \rangle \langle f |]  \nonumber  \\
& \leq & \max_{|e \rangle, |f\rangle} \sum_{j=1}^{d^2}  \frac{1}{2}( |\langle s_j |e\rangle |^4 + | \langle s_j| f \rangle |^4|)  \label{eq:sicmax1}\\
& = & \max_{|e\rangle } \sum_{j=1}^{d^2} |\langle s_j | e\rangle |^4 \label{eq:sicmax}
\eea
where the geometric mean is applied in the second inequality, $\frac{1}{n}\sum_{j=1}^n  x_j \geq (\Pi_{j=1}^n x_j )^{1/n}$. Given that $\sum_{j=1}^{d^2}\langle s_j | \rho | s_j \rangle^2/d^2=(1+\tr(\rho^2))/(d(d+1))$, as shown in \cite{rastegin}, the upper bound takes the value $\U_{d^2,d}^{(\mathrm{S})} =  2d (d+1)^{-1}$.
A similar approach yields the inequality
\bea
I_{d+1,d}^{(\mathrm{M })} (\sigma_{\mathrm{sep}}) \leq \max_{ |e\rangle} \sum_{k=1}^{d+1}\sum_{i=1}^{d} |\langle b_i^k | e\rangle |^4 \,,
\eea
for a complete set of MUBs. Using $\sum_{k=1}^{d+1}\sum_{i=1}^{d}\langle b^k_i | \rho | b^k_i \rangle^2=1+\tr(\rho^2)$, as shown in \cite{larsen}, we find $\U_{d+1,d}^{(\mathrm{M})} =  2$.
Thus, to summarize, when sets of MUBs and SICs form a quantum 2-design, we have
\bea
1 ~ \leq ~ I_{d+1,d}^{(\mathrm{M})}(\sigma_{\mathrm{sep}}) ~ \leq ~  2,  \label{eq:mubinq1}\\
\frac{d}{d+1} ~ \leq ~ I_{d^2,d}^{(\mathrm{S})}(\sigma_{\mathrm{sep}}) ~ \leq ~  \frac{2d}{d+1}.  \label{eq:sicin}
\eea
We note that these bounds have been obtained independently in Ref. \cite{ref:bae,ref:spengler,ref:li}.


\begin{thebibliography}{99}
\bibitem{footnote} Note that Alice and Bob may consider an unphysical relabeling of their basis vectors in order to optimize the correlation function. In particular, for the isotropic state  a complex conjugation in one subsystem gives the optimum.
\end{thebibliography}
\end{document}